\documentclass[12pt]{article}

\usepackage[DIV13]{typearea} % possibly not ideal for paper size US Letter
\usepackage{amsmath,amssymb,amsfonts}
\usepackage{graphicx}
\usepackage[footnotesize]{caption2}
\usepackage{bbm}
\usepackage{color}
\usepackage{braket}
\usepackage[colorlinks]{hyperref}
\usepackage[figure,table]{hypcap}

\DeclareMathOperator{\diag}{diag}

\newcommand{\MeV}{\:\text{MeV}}
\newcommand{\GeV}{\:\text{GeV}}
\newcommand{\TeV}{\:\text{TeV}}
\newcommand{\SU}{\textnormal{SU}}
\newcommand{\U}{\textnormal{U}}
\newcommand{\Mw}{M_{\textnormal{EW}}}
\newcommand{\Mp}{M_{\textnormal{P}}}
\newcommand{\Mg}{M_{\textnormal{G}}}

\newcommand{\eq}[1]{Eq.~\eqref{#1}}
\def\bea{\begin{eqnarray} }
\def\eea{ \end{eqnarray} }

\allowdisplaybreaks[1]

\begin{document}

\begin{titlepage}

\begin{center}
{\Large\bf 
Flavour issues for string-motivated heavy scalar
spectra with a low gluino mass: the $G_2$-MSSM case\\
}

\vspace{1cm}
\renewcommand{\thefootnote}{\arabic{footnote}}

\textbf{
K.~Kadota\footnote[1]{Email: \texttt{kadota@umich.edu}}$^{(a,b)}$,
G.~Kane \footnote[2]{Email: \texttt{gkane@umich.edu}}$^{(a)}$,
J.~Kersten\footnote[3]{Email: \texttt{joern.kersten@desy.de}}$^{(c)}$,
and
L.~Velasco-Sevilla\footnote[4]{Email: 
\texttt{lvelasco@fis.cinvestav.mx}}$^{(d)}$
}
\\[5mm]
\textit{\small 
$^{(a)}$
Michigan Center for Theoretical Physics, University of Michigan,\\
Ann Arbor, MI, 48104, USA\\[2mm] 
$^{(b)}$
Department of Physics, Nagoya University, Nagoya 464-8602, Japan \\[2mm]
$^{(c)}$
University of Hamburg, II.\ Institute for Theoretical Physics,\\
Luruper Chaussee 149, 22761 Hamburg, Germany\\[2mm]
$^{(d)}$
CINVESTAV-IPN, Apdo.~Postal 14-740, 07000, M\'exico D.F., M\'exico.
}
\end{center}

\vspace{1cm}

\begin{abstract}

\noindent 
In recent years it has been learned that scalar superpartner masses and
trilinear couplings should \textit{both} generically be larger than
about 20 TeV at the short distance string scale if our world is
described by a compactified string or M-theory with supersymmetry
breaking and stabilized moduli \cite{bob3}.
Here we study implications of this, somewhat generally and also in
detail for a particular realization (compactification of M-theory on a
$G_{2}$ manifold) where there is significant knowledge of the
superpotential and gauge kinetic function, and a light gluino.
In a certain sense this yields an ultraviolet completion of minimal
flavour violation.
Flavour violation stems from off-diagonal and non-universal diagonal
elements of scalar mass matrices and trilinear couplings, and from
renormalization group running. We also examine stability bounds on the
scalar potential. While heavy scalars alone do not guarantee the absence of flavour
problems, our studies show that models with heavy scalars and light
gluinos can be free from such problems.

\end{abstract}

\end{titlepage}

\setcounter{tocdepth}{3}

\section{Introduction \label{sec:intro}}

Flavour physics has usually been treated as a low-scale effective theory,
ignoring high-scale theories, except perhaps for motivating insights from short-distance physics.  It has long been thought by some that high-scale theories
with gravity-mediated supersymmetry (SUSY) breaking would typically imply too large
flavour-changing neutral current (FCNC) contributions.  Some people have argued for
heavy scalars to suppress the FCNCs (e.g., \cite{Cohen:1996vb,james}),  but these are done in the context of the low-scale effective
theory, depend on a number of detailed assumptions, and lack deeper
motivation.  Obtaining detailed flavour predictions from high-scale string
theories is difficult because it requires extensive knowledge about the
superpotential and the K\"ahler potential that is not yet available \cite{Gaillard:2005cw}.

In recent years there has been progress in constructing string/M-theories 
compactified to 4D, with broken SUSY and moduli stabilized (as is 
necessary for any theory to be a candidate to describe our world 
\cite{bob3}). Generically such theories have moduli to describe the sizes,
shapes, and other properties of the curled up dimensions. The moduli 
quanta are unstable and decay via gravitational coupling to all matter. 
If they decayed too late, the successes of big bang nucleosynthesis would be spoiled, and/or they would carry too much energy density. Their lifetime depends on their mass, so consistency with cosmology generically requires that they have masses heavier than about 20 TeV \cite{Coughlan:1983ci}.

The importance of this for flavour physics arises because in generic
string/M-theories one can show that the gravitino mass, which measures the
effects of SUSY breaking, and to which the soft-breaking Lagrangian
is proportional, must then itself be heavier than about 20 TeV\@.  Then the
supergravity theory implies that the scalar superpartner (squark
and slepton) masses, heavy Higgs masses and also the trilinear couplings must all be
larger than about 20 TeV as well.  That in turn has major effects on flavour
physics because the heavy particles and trilinears mainly decouple, though
care is needed when the off-diagonal flavour structure is included.
Studying these implications is the main goal of this paper. 

The results just stated being the generic properties of string/M-theories, if our
world is indeed described by a compactified string/M-theory with SUSY breaking and stabilized
moduli (which is crucial to define the coupling and masses needed for predictions),
it would be rather likely that the world is described by a theory with heavy scalars and
trilinears.

If we specialize to M-theory compactified on a manifold with $G_2$ symmetry,
some stronger results hold that may or may not be valid for all string
theories.  In particular in the M-theory case it has been possible to
show that the soft CP-violating phases are zero \cite{Kane:2009kv}, so there is no weak CP
problem, and also that the strong CP problem can be solved \cite{Acharya:2010zx}. For the $G_2$ compactification, once the requirement of a de Sitter vacuum and the small cosmological constant are imposed, the gaugino masses are suppressed \cite{bob5}.

With this perspective, we investigate limits from flavour and CP violation on the trilinear couplings and soft-squared scalar masses  in supersymmetric models with light gluinos and heavy scalars. Some analyses of this kind have been considered previously \cite{nima2,Giudice:2008uk,nima5}, focusing on very particular examples, perhaps with only two heavy families, and ad-hoc assumptions for Yukawa and trilinear couplings. One of the reasons for doing this is that a general analysis represents a formidable task without a priori definite information about the form of the Yukawa couplings and the
supersymmetric spectrum. However, for a particular set-up consistent with requirements of compactified string theories, we can make some
general statements and
obtain precise bounds, even though we do not yet know the precise form
of Yukawa couplings in the $G_2$-MSSM models. Our analysis is organized
as follows.

\begin{enumerate}

\item While we concentrate on the $G_2$-MSSM, we analyze three different cases:
\begin{enumerate}
\item The case where trilinear terms are proportional to Yukawa couplings at
the unification scale $\Mg$. Such a proportionality is often
used in the literature for simplicity. It arises if we have a trivial
K\"ahler potential, which respects the $G_2$ holonomy, and Yukawa
couplings that do not depend on hidden-sector fields. Due to the running of the
parameters, Yukawa and trilinear couplings are not proportional to each other at low energies. Hence, non-zero off-diagonal elements in the trilinear matrices remain even after diagonalizing the Yukawa matrices at the electroweak scale.

\item The case where trilinear terms are not proportional to Yukawa
couplings at $\Mg$, but such that the non-proportionality is determined
by real factors. Thus, complex phases at high energy enter only via the
Yukawa couplings.
This should generically happen in the context of the $G_2$-MSSM \cite{Kane:2009kv}. This case can be reproduced with a non-trivial K\"ahler potential.
We generate a series of random numbers determining the non-proportionality.

\item The case where trilinear terms are not proportional to Yukawa couplings
at $\Mg$ and where new phases appear at the high scale. This will
represent a scenario beyond the $G_2$-MSSM\@. We explore this scenario
as a contrast to the $G_2$-MSSM case so that we can determine whether or
not there could be an important impact of the phases. Again, we use
random numbers determining the non-proportionality, which are now
complex.
\end{enumerate}
We assess the impact of the trilinear terms on flavour and CP violation for each case.

\item At $\Mg$ the boundary conditions are as follows: the
Yukawa matrices of both up- and down-type quarks are non-diagonal complex
$3\times 3$ matrices.  Their diagonalizing matrices are similar to the CKM
matrix $V_\textnormal{CKM}$, so their off-diagonal elements are small,
except for the right-handed diagonalizing matrix of down quarks,
$U_R^d$, which has sizable off-diagonal elements.
For concreteness, we
use Yukawa matrices constructed in a grand unified model with a family
symmetry \cite{Kane:2005va}.  It is important to mention that we have
taken this as a definite example but this Yukawa pattern can be embedded in other contexts.  The trilinear terms are also non-diagonal complex $3\times 3$ matrices,
either (a) proportional or (b) not proportional to the Yukawa
matrices without new phases or (c) not proportional to the Yukawa matrices with new CP phases.  The
soft-squared mass matrices are proportional to the unit matrix at $\Mg$.
Recall that whenever we have a trivial K\"ahler metric, the soft-squared 
masses at that scale will be proportional to the unit matrix, because the 
same matrices diagonalizing the K\"ahler metric will diagonalize the 
soft-squared matrices. Non-trivial K\"ahler metrics could also reproduce 
diagonal soft-squared masses but lift the universality condition. 
As long as $m^2_{\tilde q_i}-m^2_{\tilde q_j}\lesssim 1.5 \, m_0^2$ at
the GUT scale, the result from this analysis will be 
valid.\footnote{If this inequality is violated, there arises too large an 
off-diagonal term in the super-CKM basis.}

\item We focus our studies on supersymmetric mass spectra featuring heavy
scalars ($m_{\tilde{q}}\gtrsim 20\TeV$) and light gauginos.
In particular, the light gluino ($m_{\tilde g}\lesssim 1\TeV$), due to
its strong interactions, can potentially play a significant role for
low-energy observables even if the scalars are heavy. For the specification
of such SUSY mass spectra, we use the $G_2$-MSSM \cite{bob5,bob4,Acharya:2008hi}
as a concrete UV-complete model, which
helps us to clarify the potential effects of the high-energy physics on the
flavour physics phenomena at the electroweak scale.
The model is based on the effective field theories arising from a class
of $\mathcal{N}=1$ fluxless compactifications of M-theory on a $G_2$
manifold.  For concreteness, we choose a set of benchmark $G_2$-MSSM
spectra that has been analyzed in \cite{Acharya:2008hi}.

\item In addition to bounds coming mainly from the kaon sector, we also
consider constraints from the stability of the scalar potential, which are
relevant for heavy spectra since they are independent of the mass scale of
the supersymmetric particles.
\end{enumerate}

In Section \ref{G2results} for completeness of this work, we summarize the defining features of  $G_2$-MSSM models.

\section{Yukawa couplings and trilinear terms}
\label{sec:YukawaTrilinear}

One goal of this analysis is to set bounds on the trilinear and soft-squared
masses.
The general relation in supergravity theories \cite{Brignole:1997dp}
between trilinear and Yukawa couplings is 
\begin{eqnarray}
a_{{\alpha }{\beta }{\gamma }} &=&\braket{ {\mathcal{F}}^m }\left[ \Braket{
\frac{\partial_m K_\text{H}}{\Mp^2}}Y_{{\alpha }{\beta }{\gamma }}+\frac{{%
\mathcal{N}}\partial Y_{{\alpha }{\beta }{\gamma }}^{\prime }}{\partial {%
\braket{h_m}}}\right]  \notag \\
&&{}-\braket{ {\mathcal{F}}^m }\left[ \Braket{\tilde{K}^{\delta\bar\rho} \,
(\partial_m \tilde K_{\bar\rho{\alpha}})}Y_{\delta {\beta }{\gamma }}+({%
\alpha }\leftrightarrow {\beta })+({\alpha }\leftrightarrow {\gamma })\right]
,  \label{eq:aContino}
\end{eqnarray}
where $\tilde{K}_{\bar{\alpha}\beta }= \frac{\partial ^{2}K}{\partial
C_{\bar{\alpha}}^{\dagger }\partial C_{\beta }}$ with
 $C_\alpha \in \{Q,u^{c\dagger},d^{c\dagger },L,e^{c\dagger },H_{u},H_{d}\}$, that is the Greek indices help to differentiate among the different chiral superfields of the theory. Greek indices with bar are related to operations on the antichiral superfields, e.g.~${C^\dagger}$.
Here $\tilde{K}^{\gamma \bar{\delta}}$
denotes the elements of the inverse matrix.
Besides, $h_{m}$ are hidden-sector fields whose $\mathcal{F}$-term
vacuum expectation values break SUSY, $K_\text{H}$ is the part of the
K\"{a}hler potential that depends only on these fields, 
$\partial_m = \partial /\partial h_{m}$ and
$\partial_{\bar{m}}^\ast = \partial/\partial h_{\bar{m}}^\ast$.
After taking the flat limit, the visible-sector superpotential has to be
rescaled as $W_{\text{O}}^{\prime }=W_{\text{O}}\ \left\langle \frac{W_{%
\text{H}}^{\ast }}{|W_{\text{H}}|}\,e^{\frac{1}{2M_{\textnormal{P}}^{2}}%
\sum_{m}|h_{m}|^{2}}\right\rangle = \mathcal{N}\,W_{\text{O}},$ where $%
W_{\text{H}}$ is the superpotential of the hidden sector and $M_{\textnormal{%
P}}$ is the reduced Planck mass. The primed quantities enter into $W_{\text{O%
}}^{\prime }$ and the unprimed ones into $W_{\text{O}}$. 
For simplicity, we assume a
trivial matter K\"{a}hler metric $\tilde{K}_{\bar{\alpha}\beta}$. In this
case the soft-squared scalar masses are proportional to the unit matrix, and
the second line in Eq.~\eqref{eq:aContino} vanishes. However, we allow the
Yukawa couplings to depend non-trivially on $h_{m}$. Consequently, the
second term in Eq.~\eqref{eq:aContino} gives a contribution to the
trilinears that is not proportional to the Yukawa matrix. In other words,
what we explore here is 
\begin{equation} \label{eq:tril_non_prop_Yuk_a}
(a^{f})_{ij}=c_{ij}^{f}A_{\tilde{f}}Y_{ij}^{f},
\end{equation}
where $i,j \in \{1,2,3\}$ are family indices, $f \in \{u,d,e\}$, and $c_{ij}^f$ are
unknown numbers determining the non-proportionality. 

It has been realized in \cite{Olive:2008vv} that only for $A_{\tilde{f}}=0$
and $m_{\tilde{f}}^{2}\propto \mathbbm{1}$ at $\Mg$ or at the
scale where the boundary conditions of the set-up are given, we can realize
at low energies, near the electroweak scale $M_{\textnormal{EW}}$, the
Minimal Flavour Violation (MFV) condition \cite{D'Ambrosio:2002ex}. 
However, even with large $A_{\tilde{f}}$, this does not imply that
FCNCs cannot be under control. In fact,
even in models with a light supersymmetric spectrum, family symmetries are a
nice way to control dangerous FCNCs \cite%
{Olive:2008vv,Abel:2001cv,Ross:2002mr,Ross:2004qn,Antusch:2007re,Antusch:2008jf,Calibbi:2009ja,Calibbi:2010rf,Kadota:2010cz,Antusch:2011sq}%
. For heavy scalar masses, one may expect that supersymmetric effects will
mostly decouple, hence ameliorating the SUSY flavour problem. 
For the concrete examples to be discussed in \S \ref{examplesection}, for
instance, FCNCs and CP violation will be suppressed because of the hierarchy
between the gaugino and the scalar masses. 
However, given the precision of observations especially in the kaon sector,
even suppressed SUSY contributions can be relevant.

\section{Most sensitive FCNC observables}
\label{sec:Observables}

The most important indirect tests that most scenarios for physics beyond
the Standard Model (SM) have to face are the electroweak precision observables, 
the anomalous magnetic moment of the muon, FCNCs, and CP violation. For the
$G_2$-MSSM examples we shall discuss in \S \ref{examplesection}, for
instance, the electroweak parameters are worked out in
such a way that contributions due to the large values of Higgs masses
involved in the theory are avoided. The Higgs sector behaves as an effective
single doublet, with one light scalar and the other mass eigenstates heavy.

In the FCNC sector the $K^0-\bar K^0$ observables $\epsilon$ and $\epsilon'$ can indeed give us a hint of ways to restrict boundary conditions of soft terms at $\Mg$.
 In this section we discuss the computation of these parameters. Recall that QCD corrections are important for these observables and therefore the different scales involved in the determination of  $\epsilon$ and $\epsilon'$ play an important role.
In \S \ref{G2results}, where we consider specific examples, we mention  other
processes as well, for example, $l_i\rightarrow l_j \gamma$, $b
\rightarrow s \gamma$, and $D^0$-$\bar D^0$ mixing, which are not constraining.

\subsection[epsilon]{$\boldsymbol\epsilon$}

The CP-violating parameter in neutral kaon mixing is defined as 
\begin{equation}  \label{epK}
\epsilon=\frac{\exp(i \pi/4)}{\sqrt{2}}\frac{\text{Im} \langle K^0|H_\text{%
eff}^{\Delta S=2}|\bar{K}^0 \rangle}{\Delta m_K}
\end{equation}
with
$\Delta m_K = 2 \, \text{Re} \langle K^0|H_\text{eff}^{\Delta S=2}|\bar{K}^0
\rangle$,
where $H_\text{eff}^{\Delta S=2}$ is the effective Hamiltonian describing
$\Delta S=2$ transitions in the $K^0$-$\bar K^0$ system. The SM prediction
and the experimental value of $\epsilon$ are \cite{Altmannshofer:2009ne}
\begin{eqnarray}
\epsilon^\text{SM} &=& (1.91\pm 0.30) \times 10^{-3},  \notag \\
|\epsilon|^\text{exp} &=& (2.228\pm 0.011) \times 10^{-3},
\label{eq:espK_SM_exp}
\end{eqnarray}
respectively.
It is well-known that gluino interactions typically give the most 
relevant SUSY
contributions to $\epsilon$ for general soft parameters. How important these
are when the scalars are heavy while the gluino remains light is an
interesting question on its own. We know that the SM and gluino-sdown
contributions to $\langle K^0|H_\text{eff}^{\Delta S=2}|\bar{K}^0 \rangle$
are proportional to $\frac{\alpha_W^2}{4 M_{\textnormal{W}}^2}[(V^*_{td}V_{ts})^2
S(x_t) + (V^*_{cd}V_{cs})^2 S(x_c) + 2(V^*_{cd}V_{cs})(V^*_{td}V_{ts})
S(x_t,x_c)] + \frac{\alpha_s^2}{4 m_{\tilde g}^2} k_{\tilde g \tilde d}
G_{\tilde g}(x_{\tilde g})$, where $x_t=m_t^2/M_{\textnormal{W}}^2$, $x_c=m_c^2/M_{%
\textnormal{W}}^2$ and $x_{\tilde g}=m_{\tilde g}^2/m_{\tilde d}^2$.%
\footnote{The values used here are those from \cite{pdg}. The functions
$S(x_i,x_j)$ are the well-known Inami-Lim functions \cite{Inami:1980fz}
entering in the SM box contributions; $S(x)$ is listed also in Appendix
\ref{sec:loop_fcts}. $G_{\tilde g}(x_{\tilde g})$ is the loop function of the box diagram involving internal gluinos and squarks and is defined in Appendix \ref{sec:loop_fcts}.}

In Figure \ref{fig:Ds_loop_gl_sm_1coups} we have plotted $m_{\tilde g}$
against $\log[F^{\tilde g}(m_{\tilde g})/F^W(M_{\textnormal{W}})]$, where
$F^W(M_\textnormal{W})=\frac{\alpha_W^2}{4 M_{\textnormal{W}}^2} S(x_t)$
and $F^{\tilde g}(m_{\tilde g})=\frac{\alpha_s^2}{4 m_{\tilde g}^2} G_{\tilde g}(x_{\tilde g})$
for two different values of the down squark mass, $m_{\tilde d}\in\left\{400, 10000\right\}\GeV$, from top to bottom of the graph.
In the models that we study here, we can never have scalar masses as
low as $400\GeV$. We show the $400\GeV$ curve to illustrate the order of
magnitude of the effective coupling entering into the CP-violating
parameter $\epsilon$ that such a scalar would produce 
and to compare with the effect of a heavy scalar with mass above $10\TeV$.
From the figure we see that even if $m_{\tilde d}=10\TeV$, a coupling
three to four orders
of magnitude bigger than the SM coupling would make the supersymmetric
contribution comparable to that of the SM\@. Such an enhancement factor is
possible due to the strong suppression of the SM contribution. For instance, 
$|\text{Im}\{(V_{td}V_{ts}^*)^2\}| \approx 2 A^4\eta \lambda^{10} \sim
10^{-7}$, where $A$, $\eta$ and $\lambda$ are the well-known Wolfenstein parameters \cite{pdg}.

\begin{figure}
\centering
\includegraphics[width=9cm]{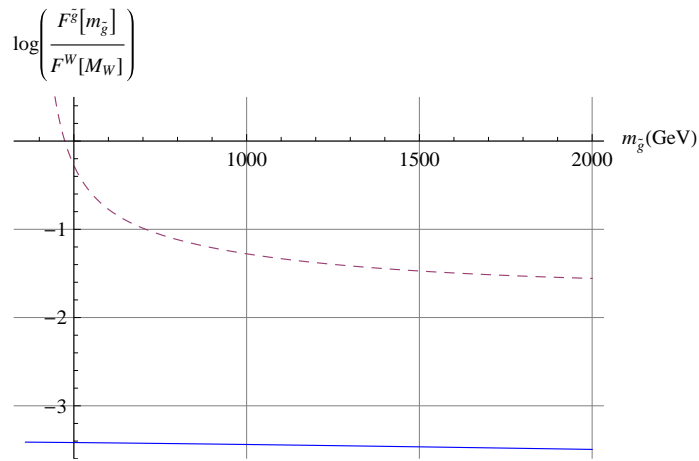}
\caption{ From top to bottom, the curves
$\log[F^{\tilde g}(m_{\tilde g})/F^W(M_{\textnormal{W}})]$ for
$m_{\tilde d}\in\{400, 10000\}\GeV$.
We can see that for $m_{\tilde d}=400\GeV$ we need a coupling a bit more
than three orders of magnitude bigger than in the SM in order to make the
SUSY contribution comparable to the SM one.}
\label{fig:Ds_loop_gl_sm_1coups}
\end{figure}

For the subject of this work, we are going to see, once we choose a  definite flavour structure at $\Mg$ encompassing non-diagonal Yukawa and trilinear couplings, even if large mixing is present, flavour-violating effects arising from  the running of soft parameters (from $\Mg$ down to $M_\textnormal{EW}$ and then to the kaon scale) are much smaller than the order of magnitude of the SM contribution and therefore flavour violation is under control in these models.

The details of the boundary conditions at $\Mg$ and the
running down to the decoupling scale of scalars are given in Appendix \ref%
{app:runn}. We follow \cite{Bertolini:1990if} to compute the Wilson
coefficients from the $\Delta S=2$ SUSY processes at the
decoupling scale of the heavy scalars. We have checked explicitly that
in our scenario the gluino indeed gives the largest supersymmetric
contribution. From the scale at which scalars decouple, $\mu_{\tilde f}$, we
then follow the effective Hamiltonian approach \cite{Contino:1998nw}.

There are two steps in the evolution from $\mu_{\tilde f}$ down to the kaon
scale. The first step is to evolve the effective Hamiltonian to the scale
where the gluinos decouple $\mu_{\tilde g}$ and the second step is to evolve
from there down to the kaon scale. The Wilson coefficients involved in
\begin{equation} \label{eq:EffHam}
\langle K^0|H_\text{eff}^{\Delta S=2}|\bar{K}^0 \rangle =
\sum_{i=1}^5 C_i \braket{O_i} + \sum_{i=1}^3 \tilde C_i \braket{\tilde O_i}
\end{equation}
at $\mu_{\tilde f}$ are 
\begin{eqnarray}
C_1 &=& -\frac{\alpha_s^2}{216 m^2_{\tilde{q}}} {(\delta^d_\text{LL})}^2_{12}
\left[ 24 x f_6(x)+66 \tilde f(x) \right],  \notag \\
C_2 &=& -\frac{\alpha_s^2}{216 m^2_{\tilde{q}}} {(\delta^d_\text{RL})}^2_{12}
\,204 x f_6(x),  \notag \\
C_3 &=& \frac{\alpha_s^2}{216 m^2_{\tilde{q}}} {(\delta^d_\text{RL})}^2_{12}
\, 36 x f_6(x),  \notag \\
C_4 &=& -\frac{\alpha_s^2}{216 m^2_{\tilde{q}}} \left[
{(\delta^d_\text{LL})}_{12} {(\delta^d_\text{RR})}_{12}
\left[504 x f_6(x)- 72 \tilde f_6(x)\right] -
{(\delta^d_\text{LR})}_{12} {(\delta^d_\text{RL})}_{12}
\, 132 \tilde f_6(x)
\right],  \notag \\
C_5 &=& -\frac{\alpha_s^2}{216 m^2_{\tilde{q}}} \left[
{(\delta^d_\text{LL})}_{12} {(\delta^d_\text{RR})}_{12}
\left[ 24 x f_6(x) + 120 \tilde f_6(x) \right] -
{(\delta^d_\text{LR})}_{12} {(\delta^d_\text{RL})}_{12}
\, 180 \tilde f_6(x)
\right],
\end{eqnarray}
where the operators are given in Appendix \ref{ssc:wilsoncoeff}. The
coefficients $\tilde C_i$ and operators $\tilde O_i$ are obtained from $C_i$
and $O_i$, respectively, by interchanging $\text{L} \leftrightarrow \text{R}$. The
functions $f_6$ and $\tilde{f}_6$ are defined in Appendix \ref{sec:loop_fcts}.
The mass-insertion parameters are defined as usual,
\begin{equation} \label{eq:DefMI}
(\delta^d_\text{XY})_{ij} =
\frac{(\hat m^2_{\tilde d\text{XY}})_{ij}^{}}%
{\sqrt{(\hat m^2_{\tilde d\text{XY}})_{ii}^{}(\hat m^2_{\tilde d\text{XY}})_{jj}^{}}},
\end{equation}
where X,Y $\in\{$L,R$\}$ and where a hat denotes a matrix in the
super-CKM (SCKM) basis \cite{Hall:1985dx}, where Yukawa couplings are diagonal.\footnote{That is, $Y^f_{\text{diag}}=\hat Y^f=U^f_\text{R}Y^f U^{f\dagger}_\text{L}$ and consequently trilinear terms are rotated as $\hat a^f=U^f_\text{R}a^f U^{f\dagger}_\text{L}$ and soft mass squared matrices as ${\hat m}^2_{\tilde f\text{LL}}=U^f_\text{L} m^2_{\tilde Q} U^{f\dagger}_\text{L}$ and ${\hat m}^2_{\tilde f\text{RR}}=U^f_\text{R} m^2_{\tilde f} U^{f\dagger}_\text{R}$.}

We take the results of \cite{Contino:1998nw} as a first approximation for
the effective Hamiltonian at the kaon scale. In particular, we use the
values for the low-energy Wilson coefficients given in that work. This
neglects the fact that in the scenario of \cite{Contino:1998nw} one sfermion
family is significantly lighter than the two heavy ones, while the models of
our interest, to be discussed with concrete examples in \S \ref%
{examplesection}, contain scalar masses of the same order of magnitude. In
order to estimate the impact of this difference, we have calculated the
running of the strong gauge coupling due to two-loop QCD corrections with
and without the contributions of the first squark family from the scale 
where the heaviest families decouple to the gluino mass scale. The
difference between the values of $g_3(m_{\tilde g})$ in the two cases is
only about 4\%, which gives us a reason to expect the change in the 
running of
the Wilson coefficients not to be dramatic either.

\subsection[epsilon'/epsilon]{$\boldsymbol{\epsilon'/\epsilon}$}

We consider here the most important contributions to $\epsilon^{\prime
}/\epsilon$,\footnote{$\epsilon^{\prime }$ being the parameter measuring the
direct CP violation in the decay amplitude of $K\rightarrow 2\pi$, $%
\epsilon^{\prime i(\delta_2-\delta_0)} \text{Re}[A_2](\text{Re} [A_0] \text{%
Im} [A_2]/\text{Re}[A_2]-\text{Im} [A_0])/(\sqrt{2} \text{Re}[A_0])$, $A_I
e^{i\delta_I}=\langle \pi \pi (I)| H^{\Delta S=1}_\text{eff}| K^0 \rangle$, $%
I=0,2 $.} namely the gluino contributions whose significance in light of the
heavy scalars is the subject of this work. They come from the chromomagnetic
penguin operators 
\begin{eqnarray}
\overline{O}_8= \frac{g_s}{8 \pi^2} m_s \bar{s}_L \sigma^{\mu \nu} t^a G^a_{\mu
\nu}d_R,\quad \overline{\tilde O}_8= \frac{g_s}{8 \pi^2} m_s \bar{s}_R
\sigma^{\mu \nu} t^a G^a_{\mu \nu}d_L,
\end{eqnarray}
where $G^a_{\mu \nu}$ is the gluon field strength. The corresponding Wilson
coefficients $C_8$ and $\tilde C_8$ are defined as in \cite{Gabbiani:1996hi}.

The direct CP violation from these operators can be estimated as \cite{hage,Gabbiani:1996hi,ey,bar5} 
\begin{equation} \label{epoe}
\text{Re}\left( \frac{\epsilon^{\prime }}{\epsilon}\right) = \frac{11\sqrt3}{64\pi} 
\frac{w}{|\epsilon|\text{Re}(A_0)} \frac{m^2_{\pi}m^2_{K}}{F_{\pi}(m_s+m_d)}
\frac{\alpha_s(m_{\tilde{g}})}{m_{\tilde{g}}}
\eta B_G \,\text{Im}\left[ x \left[
\frac{\alpha_s \pi}{m^2_{\tilde d}}\right]^{-1}
\! \left(C_8(x) - \tilde C_8(x)\right) \right],
\end{equation}
where $w=\text{Re} A_2/\text{Re} A_0=0.045$ ($A_i$ represents the amplitude for
$K\rightarrow (\pi \pi)_{I=i}$), $F_{\pi}=131\MeV$ is the pion decay constant,
$B_G$ represents the uncertainty in the hadronic matrix element calculation
for the magnetic operator between $K^0$ and the 2 pion state, $\eta$
represents the running effect from $m_{\tilde{g}}$ to $m_c$,
\begin{eqnarray}
\eta=\left( \frac{\alpha_s(m_{\tilde{g}})}{\alpha_s(m_{t})} \right)^{2/21}
\left( \frac{\alpha_s(m_{t})}{\alpha_s(m_{b})} \right)^{2/23} \left( \frac{%
\alpha_s(m_{b})}{\alpha_s(m_{c})} \right)^{2/25}.
\end{eqnarray}
The contributions from $C_8$ and $\tilde{C}_8$ coefficients can be
decomposed into the chirality changing and conserving contributions as $x %
\left[\frac{\alpha_s \pi}{m^2_{\tilde d}}\right]^{-1} \left(C_8(x) - \tilde
C_8(x)\right)=\Lambda_\text{LLRR}(x) + \Lambda_g(x)$ with 
\begin{eqnarray}
\Lambda_g(x) &=& \left[(\delta^d_\text{LR})_{12}-(\delta^d_\text{RL})_{12}\right] x
\left[-\frac{1}{3}M_1(x)-3 M_2(x) \right] ,  \notag \\
\Lambda_\text{LLRR}(x) &=&
\left[(\delta^d_\text{LL})_{12}-(\delta^d_\text{RR})_{12}\right]
\frac{m_s}{m_{\tilde g}}\ x \left[-\frac{1}{3}M_3(x)-3 M_4(x) \right],
\end{eqnarray}
where the functions $M_i$ are defined in \cite{Gabrielli:1995bd,Gabbiani:1996hi} and $x=m_{\tilde g}^2/m_{\tilde d}^2$.

The chirality-changing terms, for the models under consideration in this
letter, show up in the down sector and we shall here consider the
significant gluino contributions due to the off-diagonal
$a$-terms which can arise from the non-proportionality between Yukawa and
trilinear couplings after diagonalizing Yukawa couplings 
\begin{eqnarray}
(\delta^d_\text{LR})_{12}=\frac{a^d_{12}\langle H_d \rangle}{\hat m^2_{\tilde d\text{LR}}},\quad
(\delta^d_\text{RL})_{12}= \frac{a^d_{21}\langle H_d \rangle}{\hat m^2_{\tilde d\text{RL}}},\quad
\end{eqnarray}
$\hat m^2_{\tilde d\text{XY}}$ being the average of the two diagonal elements as in \eq{eq:DefMI} which on the other hand
can keep electric dipole moments (EDMs) sufficiently small \cite{abel,masi2,kha6}.

The contributions from $(\delta^d_\text{LL})_{12}$ and
$(\delta^d_\text{RR})_{12}$ can also be relevant if they are much bigger
than $(\delta^d_\text{LR})_{12}$ and
could even overcome the enhancement factor $m_{\tilde g}/m_s$ that multiplies
this last contribution \cite{Gabbiani:1996hi}. Those chirality-conserving
mass insertion parameters however turn out to be more stringently
constrained from $\Delta m_K$ and $\epsilon$ \cite%
{Gabrielli:1995bd,Murayama:1999yn}, and they cannot make significant
contributions to $\epsilon^{\prime }$ under those constraints from those
indirect CP violations. We hence, in the following, discuss the effects of
$(\delta^d_\text{LR,RL})_{12}$ on $\epsilon^{\prime }/\epsilon$, which can
constrain
the potential new physics effects on the flavour-changing interactions that
may stem from the non-proportionality of trilinear and Yukawa couplings.

\section{Constraints from stability of the scalar potential}
\label{ccbufb}
Before performing the numerical analysis for the flavour violation
observables, let us
briefly discuss the vacuum stability bounds which constrain the
flavour-violating trilinear soft terms by requiring the absence of charge or
color breaking (CCB) minima and directions unbounded from below (UFB)
in the scalar potential
\cite{Casas:1996de}. 
CCB and UFB constraints can become particularly important or even more
stringent than those from FCNCs for large soft SUSY breaking terms,
because the former are related to the ratio of scalar masses and trilinear
couplings while the latter tend to decrease as the scale of SUSY
breaking increases.

An undesirable deep CCB minimum appears unless the trilinear scalar
couplings satisfy
\begin{eqnarray}
|\hat a^e_{ij}|^2 &\leq& \bigl[ (\hat Y_{ii}^e)^2+(\hat Y_{jj}^e)^2 \bigr]
 \bigl[ (\hat m^2_{\tilde{e}\text{LL}})^{}_{ii} +
(\hat m^2_{\tilde{e}\text{RR}})^{}_{jj} + m^2_{H_d}+|\mu|^2 \bigr],
\\
|\hat a^d_{ij}|^2 &\leq& \bigl[ (\hat Y_{ii}^d)^2+(\hat Y_{jj}^d)^2 \bigr]
\bigl[ (\hat m^2_{\tilde{d}\text{LL}})^{}_{ii} +
(\hat m^2_{\tilde{d}\text{RR}})^{}_{jj} + m^2_{H_d}+|\mu|^2 \bigr],
\label{eq:ccbconsts} \\
|\hat a^u_{ij}|^2 &\leq& \bigl[ (\hat Y_{ii}^u)^2+(\hat Y^u_{jj})^2 \bigr]
\bigl[ (\hat m^2_{\tilde{u}\text{LL}})^{}_{ii} +
(\hat m^2_{\tilde{u}\text{RR}})^{}_{jj} + m^2_{H_u}+|\mu|^2 \bigr]
\end{eqnarray}
in the SCKM basis.
Analogously to the CCB bounds, the UFB bounds for off-diagonal trilinear
scalar couplings read\footnote{%
The simplified expression \eqref{eq:ufbLepton} is derived considering the $\mathcal{D}$-flat direction
$\alpha^2=|H_d^0|^2+|\tilde{\nu}_m|^2=|\tilde{e}_{L_i}|^2=|\tilde{e}_{R_j}|^2$
$(m\ne i,j)$ in the limit
$\alpha\gg [m^2_{H_d}+|\mu|^2-(\hat m^2_{\tilde\nu})^{}_{mm})]/[(\hat Y_{ii}^e)^2 +(\hat Y_{jj}^e)^2]$
with $\alpha^2>|H_d^0|^2$ \cite{Casas:1996de}.}
\begin{eqnarray}
|\hat a^e_{ij}|^2 &\leq& \bigl[ (\hat Y^e_{ii})^2+(\hat Y^e_{jj})^2 \bigr]
\bigl[ (\hat m^2_{\tilde{e}\text{LL}})^{}_{ii} +
(\hat m^2_{\tilde{e}\text{RR}})^{}_{jj} + (\hat m^2_{\tilde\nu})^{}_{mm} \bigr],
\label{eq:ufbLepton} \\
|\hat a^d_{ij}|^2 &\leq& \bigl[ (\hat Y^d_{ii})^2+(\hat Y^d_{jj})^2 \bigr]
\bigl[ (\hat m^2_{\tilde{d}\text{LL}})^{}_{ii} +
(\hat m^2_{\tilde{d}\text{RR}})^{}_{jj} + (\hat m^2_{\tilde\nu})^{}_{mm} \bigr],
\label{eq:ufbconsts} \\
|\hat a^u_{ij}|^2 &\leq& \bigl[ (\hat Y^u_{ii})^2+(\hat Y^u_{jj})^2 \bigr]
\bigl[ (\hat m^2_{\tilde{u}\text{LL}})^{}_{ii} +
(\hat m^2_{\tilde{u}\text{RR}})^{}_{jj} +
(\hat m^2_{\tilde{e}\text{LL}})^{}_{pp} +
(\hat m^2_{\tilde{e}\text{RR}})^{}_{qq} \bigr],
\end{eqnarray}
where $m \neq i,j$ and $p \neq q$.
While one cannot give general predictions for the values of trilinear
parameters without specifying the dependence of the K\"ahler potential and
Yukawa couplings on the hidden-sector fields as pointed out in
Eq.~(\ref{eq:aContino}), we shall restrict the range of the off-diagonal
terms $\hat a^f_{ij}$ by these CCB/UFB bounds when we perform
the numerical studies in \S \ref{G2results}.

\section{Concrete examples: $G_2$-MSSM models} \label{G2results}

\subsection{General characteristics of the $G_2$-MSSM}

Let us briefly overview the basic properties of the $G_2$-MSSM and
their origin before discussing the flavour issues.

The starting point is a compactified M-theory, which is assumed to have the MSSM embedded in the $G_2$ manifold, with no extra matter, following the work of Witten \cite{Witten:2001bf}.  The gauge group is not extended from the SM one. Supersymmetry breaking arises from the gaugino condensation mechanism, which is generic in this theory, and leads to a non-vanishing gravitino  mass. The supergravity theory then allows calculating all the soft-breaking  parameters in terms of the gravitino mass (detailed calculations in \cite{bob5}).  Then the scalar (squark and slepton and Higgs sector) masses are equal to the gravitino mass with small corrections and the trilinear factors are close to the scalar masses.  
The moduli K\"{a}hler and super-potentials of $G_{2}$-MSSM models are partially
determined \cite{vac} $G_{2}$-holonomy K\"{a}hler potentials but the
matter K\"{a}hler potentials are not \cite{Acharya:2008hi}.  In M-theory the moduli are stabilized generically because all moduli occur on an equal footing in the gauge kinetic function, and it occurs in the superpotential, so the moduli have some interactions and therefore a potential with a minimum.  Their vacuum expectation values and masses can be calculated. In the $G_2$-MSSM, both moduli K\"{a}hler and super-potentials are basic ingredients used for the stabilization of moduli. However, matter K\"{a}hler and super-potentials do not play a role in the stabilization. Although these must also respect  $G_{2}$-holonomy, the many possibilities can be reduced by studying their low-energy phenomenology.

In this respect the K\"ahler metric, of the $G_2$-MSSM considered so far, is assumed to be diagonal since the families arise at singularities on the manifold that are unlikely to overlap. However, non-trivial corrections to the off-diagonal elements of the  K\"ahler metric may appear through higher corrections in terms of hidden sector fields.  Studying effects of non-diagonal and non-universal diagonal terms phenomenologically is beyond the scope of this work, however we explore here some indirect effects by allowing deviations from the consequences of assuming a trivial K\"ahler metric, that is by studying effects of the non-proportionality of trilinear and Yukawa couplings.

In any case, the phenomenology of these models is characterized by a suppression of gaugino masses relative to the gravitino and the moduli masses. 
That scalars (squarks, sleptons, etc.) should be heavier than about $30\TeV$ is more general than the $G_2$-MSSM, depending only on the generic derivation that the moduli masses are connected to the gravitino mass, the moduli masses have a lower bound of order $30\TeV$ from robust cosmological arguments, and supergravity implies the scalar masses are closely equal to the gravitino mass.  

We are now in a position to illustrate our aforementioned analysis using
examples with a concrete UV-completion. We consider for this purpose the
$G_2$-MSSM spectra shown in Table~\ref{tab2}, which are characterized by heavy scalar masses of order the gravitino mass ($m_{3/2} \gtrsim O(10)\TeV$) and a light gluino ($m_{\tilde g} \sim 500\GeV$).

\begin{table}
\centering
\begin{tabular}{|c|c|c|c|c|c|c|c|}
\hline
Parameter\rule{0pt}{3.0ex}\rule[-1.5ex]{0pt}{0pt} & Point~1 & Point~2 & 
Point~3 & Point~4 & Point~5 & Point~6 & Point~7 \\ \hline
$m_{3/2}$ \rule{0pt}{3.0ex} & 20000 & 20000 & 20000 & 20000 & 30000 & 50000
& 30000 \\ 
$\tan\beta$ \rule{0pt}{3.0ex} & 3 & 2.65 & 2.65 & 3 & 3 & 2.5 & 3 \\ 
\hline
$\mu $ & -11943 & -13377 & -13537 & -10969 & -10490 & -34019 & +17486\\
\hline
$\mathrm{LSP\,\, type} $ & \textrm{Wino} & \textrm{Wino} & \textrm{Bino} & 
\textrm{Bino} & \textrm{Bino} & \textrm{Wino} & \textrm{Bino} \\ \hline
$m_{\tilde g}$ & 401 & 449 & 622 & 492 & 1784 & 1001 & 596.8 \\ 
$m_{\widetilde \chi_1^0}$ & 145.1 & 155.6 & 189 & 170 & 473 & 373.4 & 271 \\ 
$m_{\widetilde \chi_2^0}$ & 153 & 159 & 214.3 & 181.5 & 702.4 & 397 & 334.2
\\ 
$m_{\widetilde \chi_1^{\pm}}$ & 145.2 & 155.8 & 214.5 & 181.7 & 702.6 & 373.6
& 334.2 \\ 
$m_{\tilde d_L}$,\,$m_{\tilde s_L}$ & 19799 & 19803 & 19809 & 18785 & 21052
& 49524 & 29727 \\ 
$m_{\tilde b_1}$ & 15342 & 15250 & 15224 & 14635 & 16783 & 38473 & 23236 \\
$m_{\tilde t_1}$ & 9130 & 8779 & 8662 & 8928 & 11151 & 22887 & 14264 \\ 
$m_{\tilde d_R}$ & 19848 & 19851 & 19845 & 18832 & 21096 & 49694 & 29794 \\ 
$m_{\tilde s_R}$ & 19849 & 19851 & 19856 & 18832 & 21096 & 49695 & 29767 \\ 
$m_{\tilde t_2}$ & 15342 & 15251 & 15224 & 14635 & 16783 & 38470 & 23235 \\ 
$m_{h_0}$ & 116.4 & 114.3 & 114.6 & 116.0 & 115.9 & 115.1 & 114.6 \\
$m_{H_0}$,\,$m_{A_0}$,\,$m_{H^{\pm}}$ & 24614 & 25846 & 25943 & 23158 & 25029
& 65690 & 36623 \\ \hline
\end{tabular}
\caption{Low-scale spectra for seven benchmark $G_2$-MSSM points taken from
\protect\cite{Acharya:2008hi}.  All masses are given in GeV\@.  The other SUSY particle masses besides those
shown in this table are of order the gravitino mass. }
\label{tab2}
\end{table}

\subsection{Typical mass spectra and couplings}

The stabilization of moduli requires setting up the gravitino mass as
$m_{3/2}\in(10,100)\TeV$ and as a result gives a definitive hierarchy of
masses.

\begin{enumerate}
\item {Heavy particles:} the SUSY Higgses, the superpartners of the fermions
and the Higgs\-inos are heavy, since their masses are related to the gravitino
mass as 
\begin{eqnarray}
m^2_{\bar\alpha\beta}&=&m^2_{3/2}\delta_{\bar\alpha\beta},  \notag \\
B,\mu&\sim& m_{3/2}.
\end{eqnarray}

\item {Light particles:} gauginos and SM particles. The gauginos become
light because they are suppressed when the constraints are imposed that
require a de Sitter vacuum and a small cosmological constant \cite{bob4}.

\item {Trilinear and Yukawa couplings:} the overall scale of trilinear terms
is $A_{\tilde f} = 1.5 \, m_{3/2}$ at $\Mg$ \cite{bob5}.
So far only particular cases of matter
K\"ahler potentials have been studied. These studies have considered a
proportionality between the Yukawa couplings
and the trilinear terms.
\end{enumerate}

\subsection{Running of the $G_2$-MSSM spectra}

The running of the $G_{2}$-MSSM parameters from the scale where
$G_2$-holonomy moduli are stabilized to $M_{Z}$ has been performed by
some authors \cite{bob5,bob4,Acharya:2008hi}.  These works did not take
into account the running of $3\times 3$ Yukawa and trilinear matrices
but only the running of the third-family parameters.  As these effects
and the moderate deviations from a proportionality between trilinear and
Yukawa matrices we consider cannot have a significant influence on the
masses of the superparticles, we use the results of \cite{Acharya:2008hi}
for the mass spectra of seven $G_2$-MSSM benchmark points, as shown in
Table~\ref{tab2}.  They were calculated numerically using SOFTSUSY
\cite{Allanach:2001kg}, thus taking into account the two-loop running
and ensuring correct electroweak symmetry breaking as well as the
absence of tachyons.

In order to calculate the low-energy mass-insertion parameters, we
employ a one-loop leading-log approximation of the running of the
complex Yukawa, trilinear and soft-squared mass matrices.  Numerical
checks with SPheno 3.1.5 \cite{Porod:2003um,Porod:2011nf} and SOFTSUSY
3.2.3 \cite{Allanach:2001kg} indicate%
\footnote{Neither program is completely suited for precisely the calculation required here.}
that our approximation is rough but yields the correct order of
magnitude.  We will see that this accuracy is sufficient for the
scenario studied in this work.

\subsection{Example with hierarchical Yukawa couplings \label{sbs:hieryukcoups}}

We combine the $G_2$-MSSM spectra with the Yukawa couplings as given by the
case of Fit~4 of \cite{Kane:2005va} where we have updated the values of the
Yukawa coefficients at the GUT scale $\Mg$,
{\small
\begin{eqnarray}
&&Y^d=\frac{\sqrt{2} m_b}{v \cos\beta} 0.27\left[ 
\begin{array}{ccc}
0.0014+0.0007 i & 0.0009+0.0111 i & 0.13+0.13 i \\ 
0.0055 & 0.046+0.118 i & 0.35+0.19 i \\ 
0.0018-0.0009 i & 0.069+0.058 i & -0.90+0.08 i%
\end{array}
\right]  \notag \\
&&Y^u=\frac{\sqrt{2} m_t}{v \sin\beta} 0.53\left[ 
\begin{array}{ccc}
-1.58 \times 10^{-6}-0.000017 i & -0.000076+0.000032 i & \ 0.0020+0.0020 i
\\ 
-0.00034+0.00024 i & 0.0020+0.0002 i & 0.011+0.011 i \\ 
-0.0057-0.0024 i & 0.0044+0.0115 i & 0.70+0.71 i%
\end{array}
\right]  \notag \\
&&Y^e= \frac{\sqrt{2} m_{\tau}}{v \cos\beta} \left[ 
\begin{array}{ccc}
0.0014-0.0007 i & 0.0005-0.0056 i & 0.13-0.13 i \\ 
0.0082 & 0.023-0.059 i & 0.18-0.1 i \\ 
0.0018+0.0009 i & 0.035-0.029 i & -0.99-0.09 i%
\end{array}
\right].  \label{eq:Yuk_hier_at_MG}
\end{eqnarray}
}
As boundary conditions for the trilinear couplings, we use the relation \eqref{eq:tril_non_prop_Yuk_a} and
\begin{eqnarray}
 && (a)\ c^f_{ij} = 1,
 \label{eq:c_prop_np}\\
 && (b)\ c^f_{ij} = x^f_{ij},\ x^f_{ij} \in (0,\sqrt{2}) \text{ a random number}\ \text{and} \label{eq:c_random_np}\\
 && (c)\ c^f_{ij} = x^f_{ij} e^{i \varphi^f_{ij}}, \ x^f_{ij} \in (0,\sqrt{2}), \ \varphi^f_{ij} \in (-\pi,\pi) \text{ both random numbers},  \label{eq:c_randomnum}
\end{eqnarray}
with the exception of $c^f_{33}$, which is fixed to be $1$ in order to
preserve the aforementioned prediction for the overall scale of the
trilinear couplings.  All relations are valid at $\Mg$.
The maximum absolute value of $|c^f|=\sqrt{2}$ is chosen to ensure
that the running does not create off-diagonal elements in the
soft-squared mass matrices that are larger than the diagonal elements, as
explained in Appendix~\ref{app:runn}.

For the case (c) above,  In Table \ref{tbl:c_f_random} we show the values of the coefficients $c^d$
that have produced the maximum values of the flavour-violating parameters
$(\delta^d_\text{XY})_{12}$, which are listed in Table~\ref{tbl:deltas12xy}.
For completeness we also show the values of $c^u$. We have chosen the
matrix of coefficients $c^e = (c^d)^T$. For all SM parameters we use the values of \cite{pdg}.

\subsubsection{CP violation in the kaon sector and vacuum stability constraints}
\label{examplesection} 

\paragraph{\large{\boldmath${\protect\epsilon}$}}

In the  $G_2$-MSSM  cases the SUSY contribution to $\text{Re}\{\langle K^0|H^{\Delta S=2}_\text{eff}|\bar{K}^0\rangle\}$ is really small, therefore we can express $\protect\epsilon= \protect\epsilon^\text{SM} + \delta\epsilon^\text{SUSY}$
with $\delta\epsilon^\text{SUSY} \propto \text{Im}\{\langle K^0|H^{\Delta S=2}_\text{SUSY}|\bar{K}^0\rangle\}$.

At the scale of $10\TeV$, 
$\text{Im}(\delta^d_\text{RR})_{12}$ is the leading contribution, while the other flavour-violating parameters are at least one order of magnitude smaller. The flavour-violating parameter $(\delta^d_\text{LR})_{12}$ involves a Yukawa coupling due to the chirality flip and is therefore suppressed for very heavy
scalars.  The values of $(\delta^d_\text{RR})_{12}$ in Table~\ref{tbl:deltas12xy}
yield a contribution $\delta\epsilon^\text{SUSY} \sim 10^{-6}$, safe enough in comparison to the SM contribution and the experimental limit of order $10^{-3}$.

The parameters $\delta^u_\text{LL}$, which dominate the chargino contribution
to the kaon observables \cite{Khalil:2001wr}, are  $O(10^{-5})$. However, we have checked that this chargino contribution is still smaller than that mediated by the gluino.

It is important to mention that in the down sector only $U^d_\text{L} \sim  V_\text{CKM}$, while $U^d_\text{R}$ contains large mixings. 
Together with the relatively large difference between the masses of $\tilde b_1$
and the squarks of the first and second generation, this is responsible
for the difference of the off-diagonal elements in $\delta^d_\text{RR}$ and $\delta^d_\text{LL}$ in
this case.  We have checked the validity of the mass-insertion approximation via the vertex mixing method, analogously to the discussion for the case of $b \rightarrow s \gamma$ in section IV of \cite{Kim:1998hp}.

\begin{figure}
\centering
\includegraphics[width=11.5cm]{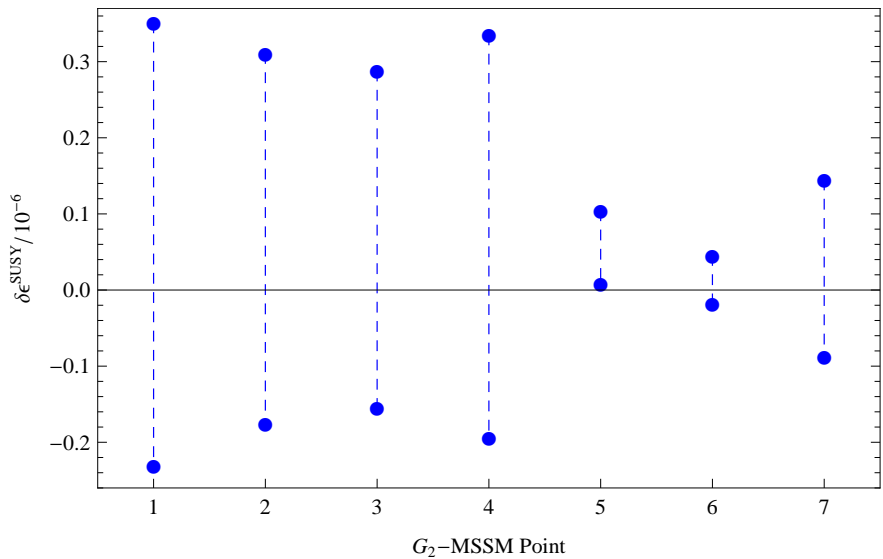}
\caption{We plot the supersymmetric contribution to $\epsilon$ for the
case that trilinear terms are not proportional to Yukawa couplings but
there are no new phases involved. If we add these values to the SM value
together with the corresponding uncertainty, we can be in agreement with
the experimental value \eqref{eq:espK_SM_exp}, since the order of
magnitude of $\delta\epsilon^\text{SUSY}$ is $O(10^{-7})$.}
\label{fig:epsk_allG2_np_Y}
\end{figure}

In Figure \ref{fig:epsk_allG2_np_Y} we plot  the predictions for
$\delta\epsilon^\text{SUSY} \approx \delta\epsilon^{\tilde g} + \delta\epsilon^{H^\pm}$  for the set-up \eqref{eq:c_random_np}, where trilinear terms are not proportional to Yukawa couplings but no new phases are involved.%
\footnote{We express $\epsilon$ as $\epsilon = \epsilon^\text{SM} + \delta\epsilon^\text{SUSY},
\quad \delta\epsilon^\text{SUSY}= \delta \epsilon^{H^{\pm}} + \delta
\epsilon^{\tilde\chi^\pm}+ \delta \epsilon^{\tilde\chi^{0}} +
\delta \epsilon^{\tilde\chi^{0} \tilde g} + \delta \epsilon^{\tilde g}$,
where $\delta\epsilon^\text{SUSY}$ is
the total SUSY contribution and the individual terms refer
to the charged Higgs, the chargino, the neutralino, the
neutralino-gluino, and the gluino contribution, respectively.}
Considering the uncertainty of $\epsilon^\text{SM}$, \eq{eq:espK_SM_exp}, it is possible to be in agreement with the experimental value $\epsilon^\text{exp}$ for all the benchmark points.\footnote{For the uncertainties in the hadronic matrix element calculations, we have used the bag parameters of \cite{Conti:1998ys}.}
In fact, the order of magnitude of the supersymmetric contribution,
$\delta\epsilon^\text{SUSY}$, is the same, that is  $O(10^{-7})$, as in the case with trilinear terms proportional to Yukawa couplings. In both cases, no points violate the CCB/UFB bounds.

\begin{figure}
\centering
\includegraphics[width=11.5cm]{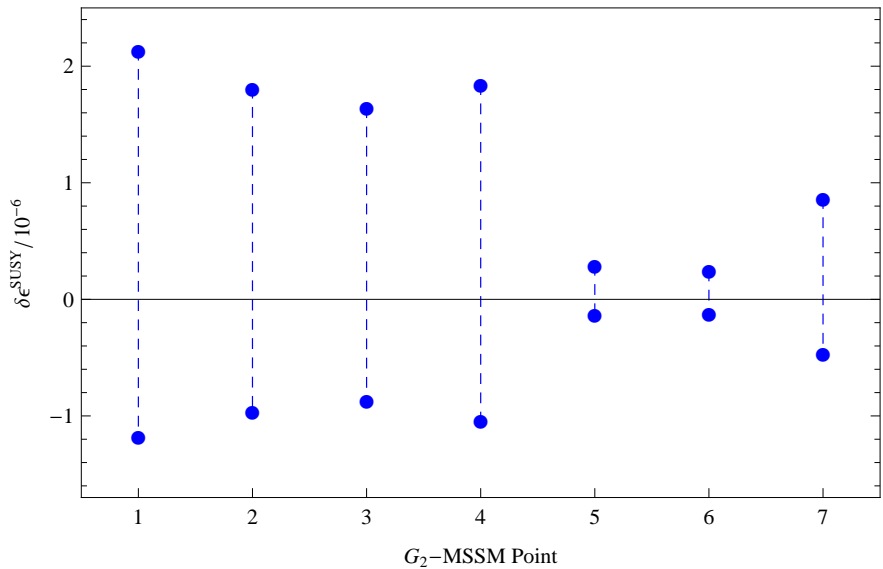}
\caption{The same as in Figure \protect\ref{fig:epsk_allG2_np_Y}, except
that in this case trilinear terms are not proportional to Yukawa
couplings and new phases appear, i.e., we have used
\eq{eq:tril_non_prop_Yuk_a} with complex coefficients \smash{$c^f_{ij}$}.
 For each $G_2$-MSSM point we  obtain values in agreement with the experimental
value $\protect\epsilon^\text{exp}$ since  all contributions to $\delta\epsilon^{SUSY}$ are of $O(10^{-6})$.}
\label{fig:epsk_allG2}
\end{figure}

Finally in Figure \ref{fig:epsk_allG2} we plot the maximum ranges of values
of $\delta\epsilon^\text{SUSY}$ 
for the case where trilinear terms are not proportional to Yukawa
couplings and where we have used Eq.~\eqref{eq:tril_non_prop_Yuk_a}
with the complex random coefficients $c^f_{ij}$ as defined in
Eq.~\eqref{eq:c_randomnum}.  We have considered only such values for the
coefficients that are allowed by the CCB/UFB constraints.  This excludes
about 10\% of the points in the random scan.
Figure \ref{fig:epsk_allG2} shows that for each $G_2$ point all the
values are in agreement with the experimental value $\epsilon^\text{exp}$ since 
all contributions to $\delta\epsilon^\text{SUSY}$ are still
at most of $O(10^{-6})$.
The benchmark points $5$--$7$ yield significantly smaller
SUSY contributions due to the larger gravitino mass and consequently
heavier scalars compared to points $1$--$4$.

\paragraph{\boldmath${\text{Re}\large{(\protect\epsilon^{\prime }/\protect\epsilon)}}$}

All kinds of mass insertions contribute to $\epsilon'$ \cite{Gabbiani:1996hi}, however those potentially large are the ones multiplied by the factor $m_{\tilde g}/m_s$, which are $\delta^d_\text{LR}$ and $\delta^d_\text{RL}$, contained in the sum of the terms $C_8 O_8+ \tilde C_8 \tilde O_8 \supset H^{\Delta S=1}_\text{SUSY} $. Due to the hierarchy of mass insertions we have found in this example, $(\delta^d_\text{RR})_{12} \gtrsim$ $(\delta^d_\text{LL})_{12} \gg $ $(\delta^d_\text{LR})_{12} \sim $ $(\delta^d_\text{RL})_{12}$, we have checked if contributions from $(\delta^d_\text{LL})_{12}$ and $(\delta^d_\text{RR})_{12}$ could play an important role.

The current experimental average of $\epsilon'/\epsilon$ from KTeV and NA48 is \cite{pdg} 
\begin{eqnarray}
\text{Re}\left( \frac{\epsilon^{\prime }}{\epsilon}\right)_\text{exp} = (1.65 \pm 0.26)
\times 10^{-3}.
\end{eqnarray}
With a conservative theoretical uncertainty, the SM contribution is
$0< \text{Re} ({\epsilon^{\prime }}/{\epsilon})_\text{SM}<3.3\times 10^{-3}$
\cite{geda}.

\begin{figure}
\label{fig:epefig_ynp} 
\centering
\includegraphics[width=11.5cm]{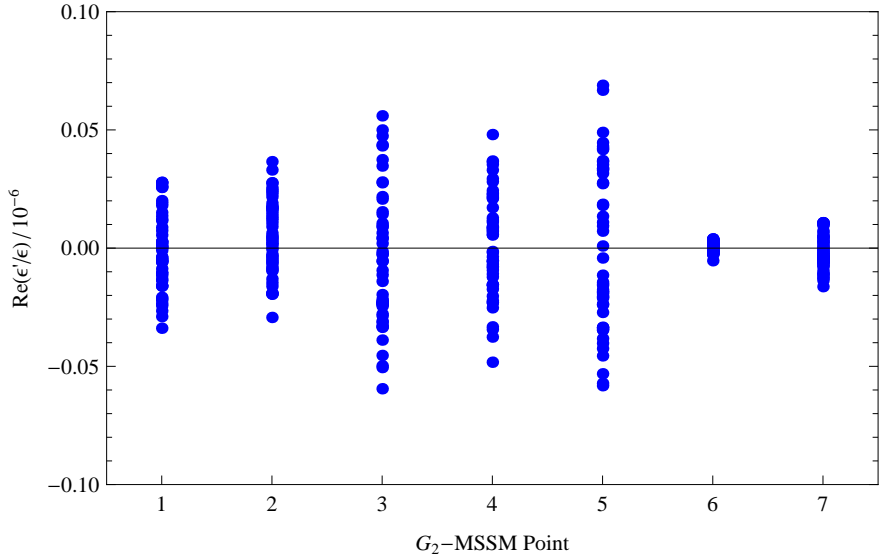}
\caption{
SUSY contribution to $\text{Re}(\epsilon'/\epsilon)$ for the case where trilinear terms
are not proportional to Yukawa couplings but there are no new phases involved, \eq{eq:c_random_np}.
It is far
smaller than the observed value $\text{Re}(\epsilon'/\epsilon) \sim 10^{-3}$.
The points correspond to different random choices of the parameters
determining the relation between trilinear and Yukawa couplings.
}
\end{figure}

\begin{figure}
\label{fig:epefig_yphases} 
\centering
\includegraphics[width=11.5cm]{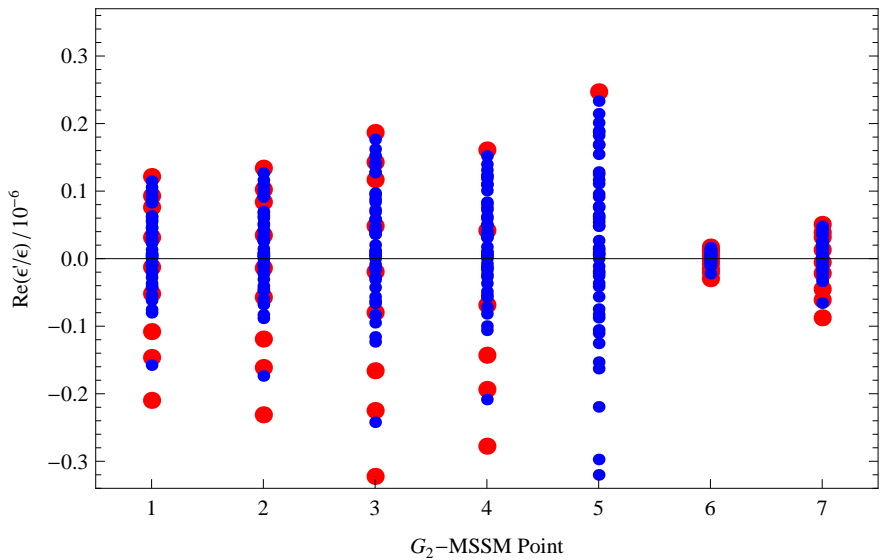}
\caption{
SUSY contribution to $\text{Re}(\epsilon'/\epsilon)$ for
the case where trilinear terms are not proportional to Yukawa couplings
and where the parameters determining the relation between trilinear and
Yukawa couplings are complex, \eq{eq:c_randomnum}.  The points
correspond to different random choices of these parameters.  The SUSY
contribution is far smaller than the observed value
$\text{Re}(\epsilon'/\epsilon) \sim 10^{-3}$.
Some of the scanned points (larger circles, in red/light shade) violate the CCB/UFB constraints.  This happens for large off-diagonal trilinear couplings, as can be seen in
Eqs.~(\ref{eq:ccbconsts},\ref{eq:ufbconsts}).
However, there are always nearby sets of parameters that give safe potentials.
}
\end{figure}

For the case of  trilinear terms proportional to Yukawa couplings,
\eq{eq:c_prop_np}, the SUSY contribution to
$\text{Re}(\protect\epsilon^{\prime }/\protect  \epsilon)$ is of the
order $10^{-9}$
for all $G_2$-MSSM points, as expected because the off-diagonal trilinear terms generated after the running are too small.
For trilinear terms not proportional to Yukawa couplings, Figure
\ref{fig:epefig_ynp} shows the values of
$\text{Re}(\epsilon'/\epsilon)$ in the case where no phases are involved,
\eq{eq:c_random_np}, while the results with new phases,
\eq{eq:c_randomnum}, are plotted in Figure~\ref{fig:epefig_yphases}.
In all cases the SUSY contribution is
smaller than $10^{-6}$ and thus negligible.

\subsubsection{Further observables \label{sec:otherprocess}}

\paragraph{Electric dipole moments.} We have discussed the effects of the off-diagonal trilinear couplings, but
there are constraints on the diagonal terms as well. For instance, the
experimental upper limit on the mercury EDM constrains the imaginary part of 
$a^{u,d}_{11}$. According to Table~3 of \cite{abe01}, 
\begin{equation}
|\text{Im}(\delta_\text{LR}^{u,d})_{11}| \lesssim 10^{-6} ,
\end{equation}
if we use $m_{\tilde q} = 20\:\text{TeV}$ and the smallest value $x =
m_{\tilde g}^2/m_{\tilde q}^2 = 0.1$ considered in \cite{abe01}. For the
smaller $x \lesssim 10^{-3}$ we encounter in the $G_2$-MSSM, the bound may
be relaxed by about an order of magnitude \cite{Kane:2009kv}.
In all the cases analyzed, we have found that
$|\text{Im}(\delta_\text{LR}^d)_{11}|$ is at most $O(10^{-8})$, while
$|\text{Im}(\delta_\text{LR}^u)_{11}|$ is at most $O(10^{-7})$.
An analogous constraint for $\text{Im}(\delta_\text{LR}^e)_{11}$ can be estimated from 
the electron EDM whose approximate contribution reads 
\cite{Masina:2002mv,Gabbiani:1996hi}
\begin{equation} 
\frac{d_e}{e}\approx \frac{\alpha_1}{4\pi}
\frac{m_{\tilde\chi^0_1}}{m^2_{\tilde{l}}} M_1(x) \,
\text{Im}(\delta_\text{LR}^e)_{11},
\end{equation}
where $x = m^2_{\tilde\chi^0_1}/m^2_{\tilde{l}}$ and the loop function $M_1(x)$ is 
given in Appendix \ref{sec:loop_fcts}. We can use this rough formula to compare with the experimental 
upper
bound of $d_e=0.07 \times 10^{-26} e\:\text{cm}$ \cite{pdg}. In our case, this limit
requires that $\text{Im}(\delta_\text{LR}^e)_{11} \lesssim 10^{-5}$. For
all the cases analyzed here, we have found that
$\text{\text{Im}}(\delta_\text{LR}^e)_{11}$ is at most $O(10^{-7})$.

\paragraph{$\mathbf{g-2}$.}

The main contributions from SUSY to the anomalous magnetic moment of the
muon, $(g-2)_\mu= 2 a_\mu$, come from the smuon-neutralino and the
sneutrino-chargino couplings \cite{Moroi:1995yh}.
The observed value of $a_\mu$ is larger than the SM prediction by
$\delta a_\mu = a^\text{exp}_\mu-a^\text{SM}_\mu= (25.9 \pm 8.1) \times 10^{-10}$
\cite{Cho:2011rk}, so if SUSY was
relevant for $a_\mu$, the muon-neutralino and the sneutrino-chargino
contributions would have to be of this order. In the case of $G_2$-MSSM spectra
with two light gauginos and a light chargino, only the diagrams involving
these particles are relevant. In this case 
\begin{eqnarray}
\delta a_\mu^\text{SUSY} &=& \delta a_\mu^{\tilde \chi^0 \tilde \mu} + \delta
a_\mu^{\tilde\chi^{\pm} \tilde \nu},  \notag \\
\delta a_\mu^{\tilde \chi^0 \tilde \mu} &\approx & -\frac{1}{16\pi^2}\frac{%
m_{\mu} m_{\tilde \chi^0_l} }{m^2_{\tilde \mu_a}} k^{a}_{\tilde\chi^0_l}\approx -2.6
\times 10^{-10} k^{a}_{\tilde\chi^0_l} ,  \notag \\
\delta a_\mu^{\tilde \chi^{\pm} \tilde \nu} &\approx & 20 \frac{1}{16\pi^2}%
\frac{m_{\mu} m_{\tilde \chi^{\pm}_{l}} }{m^2_{\tilde \nu}}
k^{a}_{\tilde\chi^\pm_l} \approx 4.4 \times 10^{-9} k^a_{\tilde\chi^\pm_l}.
\end{eqnarray}
The couplings $k^{a}_{\tilde\chi^0_l}$ are of the order
$g_1^2 v_d (-\hat{a}^{e*}_{22} + \mu \, y_\mu \tan\beta)/{((\hat{m}^2_{\tilde e \text{LL}})_{22}-(\hat{m}^2_{\tilde e \text{RR}})_{22})}$,
that is $O(g_1^2 (\delta^e_\text{LL})_{12}) \sim 10^{-5}$, while
$k^a_{\tilde\chi^\pm_1}=-K_{\tilde\chi^\pm_2} \approx -g^2_2 y_\mu \frac{v_d}{\mu}$
are at most $O(y_\mu 10^{-3})$. Both are too small to produce
a significant contribution to $(g-2)_\mu$, and thus cannot explain the deviation from the SM expectation that has been experimentally observed \cite{Bennett:2006fi}.
We stress, however, that the SM computations are not final and the predicted SM value could change~\cite{pdg}.

\paragraph{B decays} constrain the flavour-violating parameters
$\delta^d_{23}$ from their gluino-sdown, neu\-tral\-i\-no-sdown
contributions and $\delta^u_{23}$ from chargino-sup and charged
Higgs-sup contributions. We know that these processes are sensitive to
the squark mass scale and thus expected to be quite small for the models
at hand. Thus, in order to have an idea of the order of magnitude of the
decay width, $\Gamma(b\rightarrow
s\gamma)=\frac{m_b^5}{16\pi}\left|A^\gamma(\mu_b) \right|^2$, we can
estimate the contribution to the squared amplitudes. Indeed for the
cases analyzed here, the gluino-sdown contribution is the largest. This is because of the kind of Yukawa matrices we have chosen. At this scale and at leading order 
\begin{equation}
r_{C_7}=\frac{A^\gamma_{\tilde g}(\mu_b)}{A^\gamma_\text{SM}(\mu_b)} = \frac{C^{\tilde g}_7(\mu_b)}{C^\text{SM}_7(\mu_b)}
\in(-0.03,0.03),
\label{eq:bsg_gluino}
\end{equation}
where the numerical value range corresponds to the range necessary to saturate the experimental 2$\sigma$ region \cite{Barate:1998vz} and the leading-order expressions for the Wilson coefficients correspond to those of \cite{Borzumati:1999qt,Greub:1999yq,Wyler:2000mk}.
In the expression above, $\mu_b=2.6\GeV$ is the decay scale. Remember that we have to make the comparison at that scale because the gluino contribution follows a different QCD correction from $M_W$ down to $\mu_b$ \cite{Gambino:2001ew,Hurth:2003dk}. For the analysis we follow \cite{Olive:2008vv}. For all the points analyzed, we have found at most $r_{C_7}=10^{-3}$.

With a light chargino, $\tilde\chi^\pm_1$, and a light gluino, one may
wonder if the chargino-stop  and gluino-sdown loops could ever compete
significantly, in cases where the contribution could be of a concern.
In $b$ decays the leading terms in the amplitudes for these two diagrams
are proportional to
$K^u \, f_{1}(m^2_{\tilde\chi^\pm_1}/m^2_{\tilde t_1})$ and
$K^d \, f_{2}(m^2_{\tilde g}/m^2_{\tilde b_1})$.
The mixing in the $\tilde u$ and $\tilde d$ sectors is parameterized by $K^u $ and $K^d$, respectively.
The loop functions $f_1$ and $f_2$ are of course different but they 
are of similar size whenever the ratios 
${m^2_{\tilde\chi^{\pm}_1}}/{m^2_{\tilde t_1}}$ and $m^2_{\tilde
g}/m^2_{\tilde b_1}$ are comparable. Then as long as these mass ratios
are similar, a  cancellation could occur or not, depending on the
correlation of the mixing in the $\tilde u$ and $\tilde d$ sectors. In our case
this does not occur because the mixing in these sectors is quite different.

\paragraph{$\mathbf{D^0}$-$\bar{\mathbf{D}}^\mathbf{0}$ mixing} is known for setting
strong requirements on ${(\delta_\text{XY}^u)}_{ij}$. For a light
spectrum with $m_{\tilde q}$ and $m_{\tilde g}$ around $1\TeV$,
the upper limits lie between $10^{-3}$ and $10^{-1}$
\cite{Ciuchini:2007cw}. 
They are sensitive to the SUSY mass scale and become weaker for larger
$m_{\tilde q}$.  For the models discussed here, we find
${(\delta_\text{XY}^u)}_{ij} \in \left(10^{-5}, 10^{-6} \right)$, so
the SUSY contribution to $D^0$-$\bar{D}^0$ mixing is negligible.

\section{Discussion}

We have mainly focused on the effects of light gluinos on flavour- and CP-violating 
processes.
In the considered scenario with heavy scalars, the SUSY
contributions to flavour and CP observables are two to three orders of
magnitude smaller than the SM contributions.  So even if the rather
crude approximations we used underestimated the SUSY contributions by an
order of magnitude, the conclusion that they are negligible would still
hold.  Therefore, an order-of-magnitude estimate is sufficient for the
present study. We leave for future work a more general study with
improved accuracy and a set-up where we vary the mass scale of gluinos versus that of down squarks and sleptons.

It was beyond the scope of this letter to give a general detailed study of 
lepton flavour violation, partly due to the uncertainties of lepton mixing 
\cite{Paradisi:2005fk}. 
However, we have checked that the flavour-violating parameters of the $G_2$-MSSM models that we have analyzed are pretty small and safe. The orders of magnitude for all cases analyzed in \S \ref{G2results} are presented in Table~\ref{tbl:deltas-eijxy}.

In this paper, motivated by the recognition that generically compactified 
string and M-theories predict heavy sfermion masses and trilinear couplings 
($\gtrsim 20\TeV$), we have studied in detail whether or not the decoupling 
effects could leave any remaining places where phenomenological issues could arise. While no  concerns emerge, it is important to understand the effects on the relations between Yukawa couplings and trilinear terms, together with the improvement of hadronic uncertainties.  This will help to limit the sizes and phases associated to the trilinear couplings. Most interestingly, some sets of parameters could lead to CCB or UFB potentials, but nearby sets of parameters always exist which give safe potentials.  We also note that, contrary to what has been experimentally observed \cite{Bennett:2006fi}, the value we obtain for $(g-2)_\mu$ is too small to change the SM one and hence cannot provide an agreement with the experimental value at the $3\sigma$ C.L. However, the predicted SM value could change \cite{pdg} due to the many uncertainties in its calculation.

 Due to the very heavy squarks and sleptons characteristic for the studied scenario
unacceptably large flavour- or CP-violating effects
can be avoided.  Therefore, in the models suggested by the compactified
string/M-theories that predict heavy scalars and trilinears and assuming
off-diagonal elements of Yukawas and trilinears that are not unusually
large, gravity mediation of SUSY breaking does not have serious flavour and CP problems.

Let us here briefly comment on the issue of a possible tachyonic stop. While most part of flavour constraints can be relaxed by  heavy first and second
generation scalars, it has been pointed out that such heavy scalars could
drive the squared mass of the stop, $m_{\tilde t}$, negative via the renormalization group evolution unless  $m_{\tilde t} \gtrsim 7\TeV$ \cite{nima2,aga2}. In the models of our interest, however, the scalars typically have a common mass scale of order the gravitino mass, a few tens of TeV,  at the GUT scale, while the gaugino masses are typically a couple of hundred GeV at the GUT scale. Such a constraint  avoids tachyonic scalars and hence is not a concern for us in this letter.

\section*{Acknowledgments}

We particularly thank Bobby Acharya for encouragement and participating in 
the initial stages of the analysis, and we also thank 
Francesca Borzumati for clarifications on issues regarding  $b\rightarrow s \gamma$. We thank Luca Silvestrini for clarifications on the loop  functions entering into the Wilson coefficients of the $\Delta S=2$ Hamiltonian.  We thank Brent Nelson and Werner Porod for useful discussions.
J.K.~would like to thank CINVESTAV in Mexico City for hospitality during stages of this work.
This work was supported by the Michigan Center for Theoretical Physics,
and by the German Science Foundation (DFG) via the
Junior Research Group ``SUSY Phenomenology'' within the Collaborative
Research Centre 676 ``Particles, Strings and the Early Universe''.

%%%%%%%%%%%%%%%%%%%%%%%%%%%%%%%%%%%%%%%%%%%%%%%%%%%%%%%%%%%

\appendix

\begin{table}[p]
\label{tbl:deltas12xy}
\centering
\begin{tabular}{|c|c|c|c|c|}
\hline
& $|\text{Im}(\delta^d_\text{LL})_{12}^{}|\vphantom{{10^6_7}^6_j}$ & $|\text{Im}(\delta^d_\text{RL})_{12}^{}|$ &
  $|\text{Im}(\delta^d_\text{LR})_{12}^{}|$ & $|\text{Im}(\delta^d_\text{RR})_{12}^{}|$ \\ \hline
1 &
$2 \times 10^{-6}\vphantom{{10^6}^6}$ & $4 \times 10^{-7}$ & $4 \times 10^{-7}$ & $3 \times 10^{-5}$\\
2 &
$2 \times 10^{-6}$ & $4 \times 10^{-7}$ & $4 \times 10^{-7}$ & $3 \times 10^{-5}$\\
3 &
$2 \times 10^{-6}$ & $4 \times 10^{-7}$ & $4 \times 10^{-7}$ & $3 \times 10^{-5}$\\
4 &
$2 \times 10^{-6}$ & $4 \times 10^{-7}$ & $4 \times 10^{-7}$ & $3 \times 10^{-5}$\\
5 &
$1 \times 10^{-6}$ & $3 \times 10^{-7}$ & $3 \times 10^{-7}$ & $2 \times 10^{-5}$\\
6 &
$2 \times 10^{-6}$ & $2 \times 10^{-7}$ & $2 \times 10^{-7}$ & $2 \times 10^{-5}$\\
7 &
$2 \times 10^{-6}$ & $3 \times 10^{-7}$ & $3 \times 10^{-7}$ & $3 \times 10^{-5}$\\
\hline
\end{tabular}
\caption{Maximum values of $|\text{Im}{(\delta^d_\text{XY})}_{12}|$
for the scanned values of the trilinear parameters for the case of \eq{eq:c_randomnum}. }
\end{table}

\begin{table}
\centering
\begin{tabular}{|c|c|}
\hline
$c^u$ & $c^d$ \\ \hline
 {\rule{0pt}{3.5ex}\scriptsize {
$\left[
\begin{array}{ccc}
 0.28+0.12 i & -0.83-0.76 i & -0.31+1.27 i \\
 -0.86+0.88 i & -0.53-0.05 i & -0.04+0.68 i \\
 0.01-0.51 i & -0.28-0.38 i & 1
\end{array}
\right]$
}} &
{\scriptsize {$\left[
\begin{array}{ccc}
 0.68-1.09 i & 0.02-0.16 i & 1.06+0.03 i \\
 -0.1-0.57 i & -0.44-0.26 i & -0.16-0.02 i \\
 0.72-0.34 i & -0.20+0.02 i & 1
\end{array}
\right]$}}\\[2ex] \hline
\end{tabular}
\caption{Values of the coefficients $c^f_{ij}$ of trilinear couplings, \eq{eq:tril_non_prop_Yuk_a} and \eq{eq:c_randomnum}, for which the maximum values of $\delta\epsilon^\text{SUSY}$ are obtained. We have analyzed a large sample of random data and from there checked that these values give the maximum flavour violating parameters for a particular $G_2$-point or are really close to them.  The supersymmetric contributions $\protect\delta \protect\epsilon^\text{SUSY}$ lie respectively for the $G_2$-MSSM points 1 to 7 within the following ranges:
$(-1.5\times 10^{-6}\,,\, 1.8\times 10^{-6}), \
(-1.3\times 10^{-6}\,,\, 1.6\times 10^{-6}), \
(-1.1\times 10^{-6}\,,\, 1.4\times 10^{-6}), \
(-1.4\times 10^{-6}\,,\, 1.7\times 10^{-6}), \
(-1.9\times 10^{-7}\,,\, 3.2\times 10^{-7}), \
(-1.6\times 10^{-7}\,,\, 2.1\times 10^{-7}), \
(-6.1\times 10^{-7}\,,\, 7.4\times 10^{-7})$.}
\label{tbl:c_f_random}
\end{table}

\begin{table}
\label{tbl:deltas-eijxy}
\centering
\begin{tabular}{|c|c|c|c|c|}
\hline
$ij$ & $|\text{Im}(\delta^e_\text{LL})_{ij}^{}|\vphantom{{10^6_7}^6_j}$ & $|\text{Im}(\delta^e_\text{LR})_{ij}^{}|$ & $|\text{Im}(\delta^e_\text{RL})_{ij}^{}|$ & $|\text{Im}(\delta^e_\text{RR})_{ij}^{}|$ \\[0.25ex]
\hline
$12$ & $2 \times 10^{-7}\vphantom{{10^6}^6}$ & $3 \times 10^{-6}$ & $5 \times 10^{-6}$ & $8 \times 10^{-5}$\\
$13$ & $1 \times 10^{-6}$ & $2 \times 10^{-5}$ & $6 \times 10^{-5}$ & $ 2 \times 10^{-4}$ \\
$23$ & $8 \times 10^{-6}$ & $1 \times 10^{-5}$ & $8 \times 10^{-5}$ & $3 \times 10^{-4}$ \\
\hline
$ij$ & $|\text{Re}(\delta^e_\text{LL})_{ij}^{}|$ & $|\text{Re}(\delta^e_\text{LR})_{ij}^{}|$ & $|\text{Re}(\delta^e_\text{RL})_{ij}^{}|$ & $|\text{Re}(\delta^e_\text{RR})_{ij}^{}|$\\[0.25ex]
\hline
$12$ & $2 \times 10^{-7}\vphantom{{10^6}^6}$ & $3 \times 10^{-6}$ & $3 \times 10^{-6}$ & $1 \times 10^{-4}$ \\
$13$ & $1 \times10^{-6}$ & $2 \times 10^{-5}$ & $6 \times 10^{-5}$ & $3 \times 10^{-4}$\\
$23$ & $2 \times 10^{-6}$ & $1 \times 10^{-5}$ & $8 \times 10^{-5}$ & $3 \times 10^{-4}$ \\
\hline
\end{tabular}
\caption{Maximum values of the leptonic flavour-violating parameters
for the $G_2$-MSSM points analyzed.}
\end{table}

\section{Notation \label{ss:noteffham}}

\subsection{Wilson coefficients \label{ssc:wilsoncoeff}}

We follow various references \cite{Bertolini:1990if,Contino:1998nw,Bagger:1997gg} for the extraction of the effective
Hamiltonian.
The $\Delta S = 2$ operators involved in Eq.~\eqref{eq:EffHam} are 
\begin{eqnarray}
&&O_1=\bar d^\alpha \gamma_\mu P_\text{L} s^\alpha \bar d^\beta \gamma^\mu P_\text{L}
s^\beta,\quad O_2=\bar d^\alpha P_\text{L} s^\alpha \bar d^\beta P_\text{L} s^\beta, 
\notag \\
&&O_3=\bar d^\alpha P_\text{L} s^\beta \bar d^\beta P_\text{L} s^\alpha,\quad \quad \quad
O_4=\bar d^\alpha P_\text{L} s^\alpha \bar d^\beta P_\text{R} s^\beta,  \notag \\
&&O_5= \bar d^\alpha P_\text{L} s^\beta \bar d^\beta P_\text{R} s^\alpha,
\end{eqnarray}
where $P_\text{L}$ and $P_\text{R}$ are the left- and right-handed projection operators,
respectively, $\tilde O_i= O_i (\text{L}\leftrightarrow\text{R})$, and
$\braket{O_i}=\braket{\tilde O_i}$.

\subsection{Loop functions \label{sec:loop_fcts}}

We collect in this Appendix the loop functions that we have used in our analysis.
\begin{eqnarray}
S(x)&=&\frac{x (x^3- 12 x^2 + 6 x^2 \ln x + 15 x -4)}{4 (x-1)^3}  \\
G_{\tilde g}(x)&=&\frac{(x-1) x (11 x+19)-2 x (13 x+2) \ln x -1}{18 (x-1)^3 x}
\label{eq:Ax} \\
f_6(x) &=& \frac{6(1+3x)\ln x+x^3-9x^2-9x+17}{6(x-1)^5} \\
\tilde{f}_6(x) &=& \frac{6x(1+x)\ln x-x^3-9x^2+9x+1}{3(x-1)^5}
\\
M_1(x)&=&\frac{1+4x-5x^2+4x\ln x +2x^2 \ln x}{2(1-x)^4}
\end{eqnarray}

\section{Details of the running from $\Mg$ down to $\protect%
\mu_{\tilde f}$ \label{app:runn}}

We take the scalar soft squared mass matrices to be proportional to the unit
matrix at $\Mg$. Their running to $\mu_{\tilde f} \sim
m_{3/2} $, the scale at which the scalars decouple, will produce
off-diagonal entries. We require these off-diagonal elements to be
significantly smaller than the diagonal elements, since otherwise the mass-insertion approximation
would not be justified. To be concrete and conservative, let us consider $%
m_{3/2}=m_0=20\:\text{TeV}$ and demand 
\begin{equation}  \label{eq:OffDiagSmall}
(m^2_{\tilde f})_{i \neq j} \ll (10\:\text{TeV})^2
\end{equation}
at $\mu_{\tilde f}$. As we consider small values of $\tan\beta$, we can
neglect the contributions of $Y^d$ and $a^d$ to the running. Furthermore, we
consider CKM-like matrices diagonalizing $Y^u$, which implies $Y^{u\dagger}
Y^u \sim Y^u Y^{u\dagger} \sim y_t^2 \, \diag(0,0,1)$. Consequently, $Y^u$
does not affect the running of the off-diagonal elements of $m^2_{\tilde f}$.
Of course, the same is true of the terms in the renormalization group
equation (RGE) of $m^2_{\tilde f}$
that involve gauge couplings. Thus, the only relevant terms in the RGE are
those proportional to $a^{u\dagger} a^u$ and $a^u a^{u\dagger}$.
Approximating the right-hand side of the RGE by a constant value
(leading-log approximation), we then obtain at $\mu_{\tilde f}$ 
\begin{eqnarray}
|(m^2_{\tilde Q})_{i \neq j}| &\approx& 0.34 \, |(a^{u\dagger} a^u)_{ij}|, 
\label{eq:m2QRun}\\
|(m^2_{\tilde u})_{i \neq j}| &\approx& 0.68 \, |(a^u a^{u\dagger})_{ij}|, \\
|(m^2_{\tilde d})_{i \neq j}| &\approx& 0.
\end{eqnarray}
Using Eqs.~(\ref{eq:tril_non_prop_Yuk_a}, \ref{eq:c_randomnum}),
$U^u_\text{L} \sim U^u_\text{R} \sim V_\text{CKM}$
as well as $Y^u_\text{diag} \sim \diag(\lambda^8,\lambda^4,1)$,
where $\lambda \approx 0.23$ is the sine of the Cabibbo angle,
and assuming no accidental cancellations, we obtain 
\begin{equation} \label{eq:aDaggera}
|a^{u\dagger} a^u| \sim |a^u a^{u\dagger}| \sim A_{\tilde f}^2 \, x_\text{max}^2 
\begin{pmatrix}
\lambda^6 & \lambda^5 & \lambda^3 \\ 
\lambda^5 & \lambda^4 & \lambda^2 \\ 
\lambda^3 & \lambda^2 & 1
\end{pmatrix} ,
\end{equation}
where $x_\text{max}$ is the maximum value of $x^f_{ij}$.
Thus, the strongest constraint stems from
$\tilde f = \tilde u$ and $ij=23$ in Eq.~\eqref{eq:OffDiagSmall}, 
\begin{equation}
|(m^2_{\tilde u})_{23}| \sim 0.68 \, A_{\tilde u}^2 x_\text{max}^2 \lambda^2
\ll (10\TeV)^2 .
\end{equation}
With $A_{\tilde f} = 1.5 \,\frac{m_0^2}{m_{3/2}} \approx 30\TeV$,
this yields $x_\text{max} \sim 1.8$.
To be conservative, we have chosen $x_\text{max} = \sqrt{2}$ for our
numerical analysis.

\section{Comments on MFV} \label{app:MFV}

\paragraph{Trilinear terms.}

The term MFV \cite{D'Ambrosio:2002ex} refers to scenarios where \emph{all
higher-dimensional operators, constructed from SM and fields with Yukawa interactions, are invariant under CP and under the flavour group $G_F$.}
Here $G_F = \SU(3)_{q_L} \otimes \SU(3)_{u_R} \otimes \SU(3)_{d_R} \otimes \SU(3)_{l_L} \otimes \SU(3)_{e_R} \otimes \U(1)_B \otimes \U(1)_L \otimes \U(1)_Y \otimes \U(1)_{PQ} \otimes \U(1)_{e_R}$,
and the Yukawa couplings are formally regarded as auxiliary fields that
transform under $G_F$.  As a consequence,
\emph{MFV requires that the dynamics of
flavour violation is completely determined by the structure of the ordinary Yukawa couplings and in particular, all CP violation originates from the CKM phase}. 

Because of the running of all couplings of a theory, this scenario can
only be realized at one particular scale, usually a low energy scale.
Starting with parameters defined at $\Mg$, MFV can only be a good
approximation at $\Mw$, if
\begin{eqnarray}
a^f(\Mg)=Y^f(\Mg) \, A^f(\Mg),
\end{eqnarray}
where $A^f$ is a universal mass parameter for all families and kind of
fermions that is small in comparison with other soft masses of the
theory \emph{and} if Yukawa couplings are small.
This can be analyzed by studying the dependence of the RGEs of Yukawa and trilinear couplings on $Y^f$ \cite{Olive:2008vv,Ellis:2009di}.%
\footnote{One could start working in the basis where $Y^d$ is diagonal and $Y^u$ is
not. Recall that it is not possible to work in a basis where both are
diagonal precisely due to the CKM matrix.} If just the third family
Yukawa couplings are evolved, of course the size of $A^f$ does not matter because
no off-diagonal terms are produced. With a full RG evolution of
complex $3 \times 3$ Yukawa  and trilinear  matrices with small off-diagonal values, MFV can
be emulated, albeit never reproduced, for sufficiently small values of $A^{f}$ \cite{Olive:2008vv}.

\paragraph{Soft-squared masses}

The one-loop running of the soft-squared parameters $(m^2_{\tilde
f})_{ij}$ in the SCKM basis is governed by the $\beta$ functions
\begin{eqnarray}
\beta^{(1)}_{({\hat m}^2_{\tilde u_L})} &=& U^u_L(m^2_{\tilde Q}+ 2m^2_{H_u})
U^{u\dagger}_L|\hat Y^u|^2+ U^u_L(m^2_{\tilde Q}+ 2m^2_{H_d})U^{u\dagger}_LV_{%
\textnormal{CKM}} |\hat Y^d|^2V_{\textnormal{CKM}}^\dagger  \notag \\
& + &
(|\hat Y^u|^2+ V_\text{CKM}|\hat Y^d|V_\text{CKM}^\dagger) U^u_L m^2_{\tilde Q} U^{u\dagger}_L + 2
\hat Y^u (U^u_R m^2_{\tilde u} U^{u\dagger}_R)\hat Y^u  \notag \\
& + &
2V_{\textnormal{CKM}}\hat Y^d( U^d_R m^2_{\tilde d} U^{d\dagger}_R )\hat Y^d V_{%
\textnormal{CKM}}^\dagger+ 2 U^u_L a^{u\dagger} a^u U^{u\dagger}_L + 2 U^u_L
a^{d\dagger} a^d U^{u\dagger}_L
+ G_{m^2_{\tilde Q}} \mathbbm{1}
\notag \\
\beta^{(1)}_{({\hat m}^2_{\tilde f_R})}&=&
U^f_R (2 m^2_{\tilde f}+ 4 m^2_{H_f}) U^{f\dagger}_R(\hat Y^f)^2
+ 4 \hat Y^f U^f_L m^2_{\tilde Q} U^{f\dagger}_L  \notag \\
&+& 2 (\hat Y^f)^2(U^f_R m^2_{\tilde f} U^{f\dagger}_R) + 4 U^f_R (a^f a^{f\dagger})
U^{f\dagger}_R + G_{m^2_{\tilde f_R}}\mathbbm{1},
\end{eqnarray}
where $f \in \{u,d\}$ and the functions \smash{$G_{m^2_{\tilde f}}$}
contain flavour-diagonal contributions to the running involving gauge
couplings and gaugino masses. Note that at an arbitrary scale $\mu\neq
\Mg$, the terms which contain
\[
U^u_L(m^2_{\tilde Q}) U^{u\dagger}_L,\quad
U^f_R (m^2_{\tilde f}) U^{f\dagger}_R
\]
are not diagonal because of the different running of the diagonal
elements in \smash{$m^2_{\tilde Q}$ and $m^2_{\tilde f}$}.  Therefore,
off-diagonal terms will necessarily be induced.
 
Recall that even if we consider only the running of the Yukawa
couplings of the third family, this will produce a split in the masses of
$m^2_{\tilde f}$. We can always choose to go to the basis where one of
the Yukawa couplings is diagonal  {at $\Mg$, but this
does not guarantee diagonal soft mass-squared matrices in the SCKM basis}
because the fact that 
\begin{eqnarray}
(m^2_{\tilde f})_{11}= (m^2_{\tilde f})_{22} \neq (m^2_{\tilde f})_{33}
\end{eqnarray}
 {necessarily implies that not all of the matrices
$\hat m^2_{\tilde f\text{LL}} = U^{f}_{L}m^2_{\tilde Q}U^{f\dagger}_L$ and
$\hat m^2_{\tilde f\text{RR}} = U^{f}_{R}m^2_{\tilde f}U^{f\dagger}_R$ are
diagonal.} As it is known $\epsilon$ is very sensitive to this \cite{Blum:2009sk}. If the coupling 
of the particles beyond the SM was of the same
order as that of the SM particles, this would push the limit on the scale of new
physics entering into the $\Delta S=2$ processes up to 
\begin{eqnarray}
\Lambda_\text{MFV $\Delta S=2$} > 10^4 \TeV.  \label{eq:limitMFV_Deltas2}
\end{eqnarray}

In the lepton sector, we assume heavy right-handed neutrinos that decouple close to the GUT scale. Therefore only the superpartners of right- and left-handed charged leptons as well as left-handed neutrinos can induce flavour violation.   Considering the structure of fermion masses we are using, see Section \ref{sbs:hieryukcoups} and references therein, the Yukawa coupling matrix for neutrinos is the same as that for the up-quark sector at $\Mg$, therefore the flavour violation induced in this scenario is relatively small. Table~\ref{tbl:deltas-eijxy} shows the MI parameters relevant for the observables $\ell_i \rightarrow \ell_j \gamma$.

In the $G_2$-MSSM case, where all the examples that are known \cite{Kane:2009kv} correspond to the case that trilinear couplings are proportional to Yukawa couplings, we can have a theory, depending on the choice of Yukawa couplings, for which at low energy, all flavour violation present is below the experimental bounds. It is only in this sense that we can say that we have an {\it ultraviolet version of MFV} but not in the sense in which MFV is defined. For the case of the relation \eqref{eq:tril_non_prop_Yuk_a} with $c^f_{ij}$ as in \eq{eq:c_random_np}, that is $O(1)$ real random numbers between $0$ and $\sqrt{2}$, at low energies CP violating phases in addition to the CKM phase appear but also flavour violation is below experimental bounds. With $c^f_{ij}$ as in \eq{eq:c_randomnum}, i.e., with random numbers between $0$ and $\sqrt{2}$ and explicit CP phases at $\Mg$, flavour violation is more difficult to neglect but still below the experimental bounds.
This means that even with more CP phases than in the SM and large mixing present due to the choice of Yukawa couplings,
after the running to low energy we obtain a theory which satisfies flavour violation constraints. Not surprisingly, the reason are the large scalar masses. What is not a trivial result of the analysis is that still bounds on the size of trilinear terms can be obtained.

%%%%%%%%%%%%%%%%%%%%%%%%%%%%%%%%%%%%%%%%%%%%%%%%%%%%%%%%%%%%%%%%%%%%%%%%%%%%%%%%%%%%

\addcontentsline{toc}{section}{References} \frenchspacing

\providecommand{\href}[2]{#2}\begingroup\raggedright\endgroup


\begin{thebibliography}{10}

\bibitem{bob3}
B.~S. Acharya, G.~Kane, and E.~Kuflik, ``{String Theories with Moduli
  Stabilization Imply Non-Thermal Cosmological History, and Particular Dark
  Matter}'', \href{http://arxiv.org/abs/1006.3272}{{\tt arXiv:1006.3272
  [hep-ph]}}.

\bibitem{Cohen:1996vb}
A.~G. Cohen, D.~Kaplan, and A.~Nelson, ``{The more Minimal Supersymmetric
  Standard Model}'',
  \href{http://dx.doi.org/10.1016/S0370-2693(96)01183-5}{{\em Phys. Lett.} {\bf
  B388} (1996)  588--598}, \href{http://arxiv.org/abs/hep-ph/9607394}{{\tt
  arXiv:hep-ph/9607394}}.

\bibitem{james}
J.~D. Wells, ``{PeV-scale supersymmetry}'',
  \href{http://dx.doi.org/10.1103/PhysRevD.71.015013}{{\em Phys. Rev.} {\bf
  D71} (2005)  015013}, \href{http://arxiv.org/abs/hep-ph/0411041}{{\tt
  arXiv:hep-ph/0411041}}.

\bibitem{Gaillard:2005cw}
M.~K. Gaillard and B.~D. Nelson, ``{On quadratic divergences in supergravity,
  vacuum energy and the supersymmetric flavor problem}'',
  \href{http://dx.doi.org/10.1016/j.nuclphysb.2006.05.035}{{\em Nucl. Phys.}
  {\bf B751} (2006)  75--107}, \href{http://arxiv.org/abs/hep-ph/0511234}{{\tt
  arXiv:hep-ph/0511234}}.

\bibitem{Coughlan:1983ci}
G.~Coughlan, W.~Fischler, E.~W. Kolb, S.~Raby, and G.~G. Ross, ``{Cosmological
  Problems for the Polonyi Potential}'',
  \href{http://dx.doi.org/10.1016/0370-2693(83)91091-2}{{\em Phys. Lett.} {\bf
  B131} (1983)  59}.

\bibitem{Kane:2009kv}
G.~Kane, P.~Kumar, and J.~Shao, ``{CP-violating phases in M theory and
  implications for electric dipole moments}'',
  \href{http://dx.doi.org/10.1103/PhysRevD.82.055005}{{\em Phys. Rev.} {\bf
  D82} (2010)  055005},
\href{http://arxiv.org/abs/0905.2986}{{\tt arXiv:0905.2986 [hep-ph]}}.
%%CITATION = 0905.2986;%%.

\bibitem{Acharya:2010zx}
B.~S. Acharya, K.~Bobkov, and P.~Kumar, ``{An M Theory Solution to the Strong
  CP Problem and Constraints on the Axiverse}'',
  \href{http://dx.doi.org/10.1007/JHEP11(2010)105}{{\em JHEP} {\bf 1011} (2010)
   105}, \href{http://arxiv.org/abs/1004.5138}{{\tt arXiv:1004.5138 [hep-th]}}.

\bibitem{bob5}
B.~S. Acharya, K.~Bobkov, G.~L. Kane, P.~Kumar, and J.~Shao, ``{Explaining the
  Electroweak Scale and Stabilizing Moduli in M Theory}'',
  \href{http://dx.doi.org/10.1103/PhysRevD.76.126010}{{\em Phys. Rev.} {\bf
  D76} (2007)  126010}, \href{http://arxiv.org/abs/hep-th/0701034}{{\tt
  arXiv:hep-th/0701034}}.

\bibitem{nima2}
N.~Arkani-Hamed and H.~Murayama, ``{Can the supersymmetric flavor problem
  decouple?}'', \href{http://dx.doi.org/10.1103/PhysRevD.56.R6733}{{\em Phys.
  Rev.} {\bf D56} (1997)  6733--6737},
  \href{http://arxiv.org/abs/hep-ph/9703259}{{\tt arXiv:hep-ph/9703259}}.

\bibitem{Giudice:2008uk}
G.~F. Giudice, M.~Nardecchia, and A.~Romanino, ``{Hierarchical Soft Terms and
  Flavor Physics}'',
  \href{http://dx.doi.org/10.1016/j.nuclphysb.2008.12.030}{{\em Nucl. Phys.}
  {\bf B813} (2009)  156--173}, \href{http://arxiv.org/abs/0812.3610}{{\tt
  arXiv:0812.3610 [hep-ph]}}.

\bibitem{nima5}
N.~Arkani-Hamed, S.~Dimopoulos, G.~Giudice, and A.~Romanino, ``{Aspects of
  split supersymmetry}'',
  \href{http://dx.doi.org/10.1016/j.nuclphysb.2004.12.026}{{\em Nucl. Phys.}
  {\bf B709} (2005)  3--46}, \href{http://arxiv.org/abs/hep-ph/0409232}{{\tt
  arXiv:hep-ph/0409232}}.

\bibitem{Kane:2005va}
G.~Kane, S.~King, I.~Peddie, and L.~Velasco-Sevilla, ``{Study of theory and
  phenomenology of some classes of family symmetry and unification models}'',
  \href{http://dx.doi.org/10.1088/1126-6708/2005/08/083}{{\em JHEP} {\bf 0508}
  (2005)  083}, \href{http://arxiv.org/abs/hep-ph/0504038}{{\tt
  arXiv:hep-ph/0504038}}.

\bibitem{bob4}
B.~S. Acharya, K.~Bobkov, G.~L. Kane, J.~Shao, and P.~Kumar, ``{$G_2$-MSSM: An
  $M$ theory motivated model of particle physics}'',
  \href{http://dx.doi.org/10.1103/PhysRevD.78.065038}{{\em Phys. Rev.} {\bf
  D78} (2008)  065038}, \href{http://arxiv.org/abs/0801.0478}{{\tt
  arXiv:0801.0478 [hep-ph]}}.

\bibitem{Acharya:2008hi}
B.~S. Acharya and K.~Bobkov, ``{Kahler Independence of the $G_2$-MSSM}'',
  \href{http://dx.doi.org/10.1007/JHEP09(2010)001}{{\em JHEP} {\bf 09} (2010)
  001},
\href{http://arxiv.org/abs/0810.3285}{{\tt arXiv:0810.3285 [hep-th]}}.
%%CITATION = 0810.3285;%%.

\bibitem{Brignole:1997dp}
A.~Brignole, L.~E. Iba\~nez, and C.~Mu\~noz, ``{Soft supersymmetry-breaking
  terms from supergravity and superstring models}'',
\href{http://arxiv.org/abs/hep-ph/9707209}{{\tt arXiv:hep-ph/9707209}}.
%%CITATION = HEP-PH/9707209;%%.

\bibitem{Olive:2008vv}
K.~A. Olive and L.~Velasco-Sevilla, ``{Constraints on Supersymmetric Flavour
  Models from $b \to s \gamma$}'',
  \href{http://dx.doi.org/10.1088/1126-6708/2008/05/052}{{\em JHEP} {\bf 05}
  (2008)  052},
\href{http://arxiv.org/abs/0801.0428}{{\tt arXiv:0801.0428 [hep-ph]}}.
%%CITATION = 0801.0428;%%.

\bibitem{D'Ambrosio:2002ex}
G.~D'Ambrosio, G.~Giudice, G.~Isidori, and A.~Strumia, ``{Minimal flavor
  violation: An Effective field theory approach}'',
  \href{http://dx.doi.org/10.1016/S0550-3213(02)00836-2}{{\em Nucl. Phys.} {\bf
  B645} (2002)  155--187}, \href{http://arxiv.org/abs/hep-ph/0207036}{{\tt
  arXiv:hep-ph/0207036}}.

\bibitem{Abel:2001cv}
S.~Abel, S.~Khalil, and O.~Lebedev, ``{Additional stringy sources for electric
  dipole moments}'',
  \href{http://dx.doi.org/10.1103/PhysRevLett.89.121601}{{\em Phys. Rev. Lett.}
  {\bf 89} (2002)  121601},
\href{http://arxiv.org/abs/hep-ph/0112260}{{\tt arXiv:hep-ph/0112260}}.
%%CITATION = HEP-PH/0112260;%%.

\bibitem{Ross:2002mr}
G.~G. Ross and O.~Vives, ``{Yukawa structure, flavor changing, and CP violation
  in supergravity}'', \href{http://dx.doi.org/10.1103/PhysRevD.67.095013}{{\em
  Phys. Rev.} {\bf D67} (2003)  095013},
\href{http://arxiv.org/abs/hep-ph/0211279}{{\tt arXiv:hep-ph/0211279}}.
%%CITATION = HEP-PH/0211279;%%.

\bibitem{Ross:2004qn}
G.~G. Ross, L.~Velasco-Sevilla, and O.~Vives, ``{Spontaneous CP violation and
  non-Abelian family symmetry in SUSY}'',
  \href{http://dx.doi.org/10.1016/j.nuclphysb.2004.05.020}{{\em Nucl. Phys.}
  {\bf B692} (2004)  50--82},
\href{http://arxiv.org/abs/hep-ph/0401064}{{\tt arXiv:hep-ph/0401064}}.
%%CITATION = HEP-PH/0401064;%%.

\bibitem{Antusch:2007re}
S.~Antusch, S.~F. King, and M.~Malinsk\'y, ``{Solving the SUSY Flavour and CP
  Problems with SU(3) Family Symmetry}'',
  \href{http://dx.doi.org/10.1088/1126-6708/2008/06/068}{{\em JHEP} {\bf 06}
  (2008)  068},
\href{http://arxiv.org/abs/0708.1282}{{\tt arXiv:0708.1282 [hep-ph]}}.
%%CITATION = 0708.1282;%%.

\bibitem{Antusch:2008jf}
S.~Antusch, S.~F. King, M.~Malinsk\'y, and G.~G. Ross, ``{Solving the SUSY
  Flavour and CP Problems with Non-Abelian Family Symmetry and Supergravity}'',
  \href{http://dx.doi.org/10.1016/j.physletb.2008.11.020}{{\em Phys. Lett.}
  {\bf B670} (2009)  383--389}, \href{http://arxiv.org/abs/0807.5047}{{\tt
  arXiv:0807.5047 [hep-ph]}}.

\bibitem{Calibbi:2009ja}
L.~Calibbi, J.~Jones-Perez, A.~Masiero, J.-h. Park, W.~Porod, and O.~Vives,
  ``{FCNC and CP Violation Observables in a SU(3)-flavoured MSSM}'',
  \href{http://dx.doi.org/10.1016/j.nuclphysb.2009.12.029}{{\em Nucl. Phys.}
  {\bf B831} (2010)  26--71}, \href{http://arxiv.org/abs/0907.4069}{{\tt
  arXiv:0907.4069 [hep-ph]}}.

\bibitem{Calibbi:2010rf}
L.~Calibbi, E.~J. Chun, and L.~Velasco-Sevilla, ``{Bridging flavour violation
  and leptogenesis in SU(3) family models}'',
  \href{http://dx.doi.org/10.1007/JHEP11(2010)090}{{\em JHEP} {\bf 1011} (2010)
   090}, \href{http://arxiv.org/abs/1005.5563}{{\tt arXiv:1005.5563 [hep-ph]}}.

\bibitem{Kadota:2010cz}
K.~Kadota, J.~Kersten, and L.~Velasco-Sevilla, ``{Supersymmetric Musings on the
  Predictivity of Family Symmetries}'',
  \href{http://dx.doi.org/10.1103/PhysRevD.82.085022}{{\em Phys. Rev.} {\bf
  D82} (2010)  085022}, \href{http://arxiv.org/abs/1007.1532}{{\tt
  arXiv:1007.1532 [hep-ph]}}.

\bibitem{Antusch:2011sq}
S.~Antusch, L.~Calibbi, V.~Maurer, and M.~Spinrath, ``{From Flavour to SUSY
  Flavour Models}'', \href{http://arxiv.org/abs/1104.3040}{{\tt arXiv:1104.3040
  [hep-ph]}}.

\bibitem{Altmannshofer:2009ne}
W.~Altmannshofer, A.~J. Buras, S.~Gori, P.~Paradisi, and D.~M. Straub,
  ``{Anatomy and Phenomenology of FCNC and CPV Effects in SUSY Theories}'',
  \href{http://dx.doi.org/10.1016/j.nuclphysb.2009.12.019}{{\em Nucl. Phys.}
  {\bf B830} (2010)  17--94},
\href{http://arxiv.org/abs/0909.1333}{{\tt arXiv:0909.1333 [hep-ph]}}.
%%CITATION = 0909.1333;%%.

\bibitem{pdg}
Particle Data Group, K.~Nakamura {\em et al.}, ``{Review of particle
  physics}'', \href{http://dx.doi.org/10.1088/0954-3899/37/7A/075021}{{\em J.
  Phys.} {\bf G37} (2010)  075021}.

\bibitem{Inami:1980fz}
T.~Inami and C.~Lim, ``{Effects of Superheavy Quarks and Leptons in Low-Energy
  Weak Processes $K_L \to \mu \bar\mu$, $K^+ \to \pi^+ \nu \bar\nu$ and $K^0
  \leftrightarrow \bar K^0$}'',
  \href{http://dx.doi.org/10.1143/PTP.65.297}{{\em Prog. Theor. Phys.} {\bf 65}
  (1981)  297}.

\bibitem{Bertolini:1990if}
S.~Bertolini, F.~Borzumati, A.~Masiero, and G.~Ridolfi, ``{Effects of
  supergravity induced electroweak breaking on rare $B$ decays and mixings}'',
\href{http://dx.doi.org/10.1016/0550-3213(91)90320-W}{{\em Nucl. Phys.} {\bf
  B353} (1991)  591--649}.
%%CITATION = NUPHA,B353,591;%%.

\bibitem{Contino:1998nw}
R.~Contino and I.~Scimemi, ``{The supersymmetric flavor problem for heavy
  first-two generation scalars at next-to-leading order}'',
  \href{http://dx.doi.org/10.1007/s100520050597}{{\em Eur. Phys. J.} {\bf C10}
  (1999)  347--356},
\href{http://arxiv.org/abs/hep-ph/9809437}{{\tt arXiv:hep-ph/9809437}}.
%%CITATION = HEP-PH/9809437;%%.

\bibitem{Hall:1985dx}
L.~J. Hall, V.~A. Kostelecky, and S.~Raby, ``{New Flavor Violations in
  Supergravity Models}'',
\href{http://dx.doi.org/10.1016/0550-3213(86)90397-4}{{\em Nucl. Phys.} {\bf
  B267} (1986)  415}.
%%CITATION = NUPHA,B267,415;%%.

\bibitem{Gabbiani:1996hi}
F.~Gabbiani, E.~Gabrielli, A.~Masiero, and L.~Silvestrini, ``{A complete
  analysis of FCNC and CP constraints in general SUSY extensions of the
  standard model}'', \href{http://dx.doi.org/10.1016/0550-3213(96)00390-2}{{\em
  Nucl. Phys.} {\bf B477} (1996)  321--352},
\href{http://arxiv.org/abs/hep-ph/9604387}{{\tt arXiv:hep-ph/9604387}}.
%%CITATION = HEP-PH/9604387;%%.

\bibitem{hage}
J.~S. Hagelin, S.~Kelley, and T.~Tanaka, ``{Supersymmetric flavor changing
  neutral currents: Exact amplitudes and phenomenological analysis}'',
  \href{http://dx.doi.org/10.1016/0550-3213(94)90113-9}{{\em Nucl. Phys.} {\bf
  B415} (1994)  293--331}.

\bibitem{ey}
G.~Eyal, A.~Masiero, Y.~Nir, and L.~Silvestrini, ``{Probing supersymmetric
  flavor models with $\epsilon'/\epsilon$}'',
  \href{http://dx.doi.org/10.1088/1126-6708/1999/11/032}{{\em JHEP} {\bf 9911}
  (1999)  032}, \href{http://arxiv.org/abs/hep-ph/9908382}{{\tt
  arXiv:hep-ph/9908382}}.

\bibitem{bar5}
R.~Barbieri, R.~Contino, and A.~Strumia, ``{$\epsilon'$ from supersymmetry with
  nonuniversal $A$ terms?}'',
  \href{http://dx.doi.org/10.1016/S0550-3213(99)00747-6}{{\em Nucl. Phys.} {\bf
  B578} (2000)  153--162}, \href{http://arxiv.org/abs/hep-ph/9908255}{{\tt
  arXiv:hep-ph/9908255}}.

\bibitem{Gabrielli:1995bd}
E.~Gabrielli, A.~Masiero, and L.~Silvestrini, ``{Flavor changing neutral
  currents and CP violating processes in generalized supersymmetric
  theories}'', \href{http://dx.doi.org/10.1016/0370-2693(96)00158-X}{{\em Phys.
  Lett.} {\bf B374} (1996)  80--86},
  \href{http://arxiv.org/abs/hep-ph/9509379}{{\tt arXiv:hep-ph/9509379}}.

\bibitem{abel}
S.~Abel and J.~Frere, ``{Could the MSSM have no CP violation in the CKM
  matrix?}'', \href{http://dx.doi.org/10.1103/PhysRevD.55.1623}{{\em Phys.
  Rev.} {\bf D55} (1997)  1623--1629},
  \href{http://arxiv.org/abs/hep-ph/9608251}{{\tt arXiv:hep-ph/9608251}}.

\bibitem{masi2}
A.~Masiero and H.~Murayama, ``{Can $\epsilon'/\epsilon$ be Supersymmetric?}'',
  \href{http://dx.doi.org/10.1103/PhysRevLett.83.907}{{\em Phys. Rev. Lett.}
  {\bf 83} (1999)  907--910}, \href{http://arxiv.org/abs/hep-ph/9903363}{{\tt
  arXiv:hep-ph/9903363}}.

\bibitem{kha6}
S.~Khalil, T.~Kobayashi, and O.~Vives, ``{EDM-free supersymmetric CP violation
  with non-universal soft terms}'',
  \href{http://dx.doi.org/10.1016/S0550-3213(00)00239-X}{{\em Nucl. Phys.} {\bf
  B580} (2000)  275--288}, \href{http://arxiv.org/abs/hep-ph/0003086}{{\tt
  arXiv:hep-ph/0003086}}.

\bibitem{Murayama:1999yn}
H.~Murayama, ``{Can $\epsilon'/\epsilon$ be supersymmetric?}'',
  \href{http://arxiv.org/abs/hep-ph/9908442}{{\tt arXiv:hep-ph/9908442}}.

\bibitem{Casas:1996de}
J.~Casas and S.~Dimopoulos, ``{Stability bounds on flavor violating trilinear
  soft terms in the MSSM}'',
  \href{http://dx.doi.org/10.1016/0370-2693(96)01000-3}{{\em Phys. Lett.} {\bf
  B387} (1996)  107--112}, \href{http://arxiv.org/abs/hep-ph/9606237}{{\tt
  arXiv:hep-ph/9606237}}.

\bibitem{Witten:2001bf}
E.~Witten, ``{Deconstruction, $G_2$ holonomy, and doublet-triplet splitting}'',
  \href{http://arxiv.org/abs/hep-ph/0201018}{{\tt arXiv:hep-ph/0201018}}.

\bibitem{vac}
B.~S. Acharya, F.~Denef, and R.~Valandro, ``{Statistics of M theory vacua}'',
  \href{http://dx.doi.org/10.1088/1126-6708/2005/06/056}{{\em JHEP} {\bf 0506}
  (2005)  056}, \href{http://arxiv.org/abs/hep-th/0502060}{{\tt
  arXiv:hep-th/0502060}}.

\bibitem{Allanach:2001kg}
B.~C. Allanach, ``{SOFTSUSY: a program for calculating supersymmetric
  spectra}'', \href{http://dx.doi.org/10.1016/S0010-4655(01)00460-X}{{\em
  Comput. Phys. Commun.} {\bf 143} (2002)  305--331},
\href{http://arxiv.org/abs/hep-ph/0104145}{{\tt arXiv:hep-ph/0104145}}.
%%CITATION = HEP-PH/0104145;%%.

\bibitem{Porod:2003um}
W.~Porod, ``{SPheno, a program for calculating supersymmetric spectra, SUSY
  particle decays and SUSY particle production at $e^+ e^-$ colliders}'',
  \href{http://dx.doi.org/10.1016/S0010-4655(03)00222-4}{{\em Comput. Phys.
  Commun.} {\bf 153} (2003)  275--315},
\href{http://arxiv.org/abs/hep-ph/0301101}{{\tt arXiv:hep-ph/0301101}}.
%%CITATION = HEP-PH/0301101;%%.

\bibitem{Porod:2011nf}
W.~Porod and F.~Staub, ``{SPheno 3.1: Extensions including flavour, CP-phases
  and models beyond the MSSM}'',
\href{http://arxiv.org/abs/1104.1573}{{\tt arXiv:1104.1573 [hep-ph]}}.
%%CITATION = ARXIV:1104.1573;%%.

\bibitem{Khalil:2001wr}
S.~Khalil and O.~Lebedev, ``{Chargino contributions to $\epsilon$ and
  $\epsilon'$}'', \href{http://dx.doi.org/10.1016/S0370-2693(01)00880-2}{{\em
  Phys. Lett.} {\bf B515} (2001)  387--394},
\href{http://arxiv.org/abs/hep-ph/0106023}{{\tt arXiv:hep-ph/0106023}}.
%%CITATION = HEP-PH/0106023;%%.

\bibitem{Kim:1998hp}
Y.~G. Kim, P.~Ko, and J.~S. Lee, ``{Possible new-physics signals in $b \to s
  \gamma$ and $b \to s l^+ l^-$}'',
  \href{http://dx.doi.org/10.1016/S0550-3213(99)00048-6}{{\em Nucl. Phys.} {\bf
  B544} (1999)  64--88}, \href{http://arxiv.org/abs/hep-ph/9810336}{{\tt
  arXiv:hep-ph/9810336}}.

\bibitem{Conti:1998ys}
L.~Conti, C.~Allton, A.~Donini, V.~Gimenez, L.~Giusti, {\em et al.},
  ``{$B$-parameters for $\Delta S = 2$ Supersymmetric Operators}'',
  \href{http://dx.doi.org/10.1016/S0920-5632(99)85059-2}{{\em Nucl. Phys. Proc.
  Suppl.} {\bf 73} (1999)  315--317},
  \href{http://arxiv.org/abs/hep-lat/9809162}{{\tt arXiv:hep-lat/9809162}}.

\bibitem{geda}
O.~Gedalia, G.~Isidori, and G.~Perez, ``{Combining Direct and Indirect Kaon CP
  Violation to Constrain the Warped KK Scale}'',
  \href{http://dx.doi.org/10.1016/j.physletb.2009.10.097}{{\em Phys. Lett.}
  {\bf B682} (2009)  200--206}, \href{http://arxiv.org/abs/0905.3264}{{\tt
  arXiv:0905.3264 [hep-ph]}}.

\bibitem{abe01}
S.~Abel, S.~Khalil, and O.~Lebedev, ``{EDM constraints in supersymmetric
  theories}'', \href{http://dx.doi.org/10.1016/S0550-3213(01)00233-4}{{\em
  Nucl. Phys.} {\bf B606} (2001)  151--182},
  \href{http://arxiv.org/abs/hep-ph/0103320}{{\tt arXiv:hep-ph/0103320}}.

\bibitem{Masina:2002mv}
I.~Masina and C.~A. Savoy, ``{Sleptonarium: Constraints on the CP and flavor
  pattern of scalar lepton masses}'',
  \href{http://dx.doi.org/10.1016/S0550-3213(03)00294-3}{{\em Nucl. Phys.} {\bf
  B661} (2003)  365--393}, \href{http://arxiv.org/abs/hep-ph/0211283}{{\tt
  arXiv:hep-ph/0211283}}.

\bibitem{Moroi:1995yh}
T.~Moroi, ``{The Muon anomalous magnetic dipole moment in the minimal
  supersymmetric standard model}'',
  \href{http://dx.doi.org/10.1103/PhysRevD.53.6565,
  10.1103/PhysRevD.56.4424}{{\em Phys. Rev.} {\bf D53} (1996)  6565--6575},
  \href{http://arxiv.org/abs/hep-ph/9512396}{{\tt arXiv:hep-ph/9512396}}.

\bibitem{Cho:2011rk}
G.-C. Cho, K.~Hagiwara, Y.~Matsumoto, and D.~Nomura, ``{The MSSM confronts the
  precision electroweak data and the muon $g-2$}'',
  \href{http://arxiv.org/abs/1104.1769}{{\tt arXiv:1104.1769 [hep-ph]}}.

\bibitem{Bennett:2006fi}
Muon $(g-2)$ Collaboration, G.~Bennett {\em et al.}, ``{Final Report of the
  Muon E821 Anomalous Magnetic Moment Measurement at BNL}'',
  \href{http://dx.doi.org/10.1103/PhysRevD.73.072003}{{\em Phys. Rev.} {\bf
  D73} (2006)  072003}, \href{http://arxiv.org/abs/hep-ex/0602035}{{\tt
  arXiv:hep-ex/0602035}}.

\bibitem{Barate:1998vz}
ALEPH Collaboration, R.~Barate {\em et al.}, ``{A measurement of the inclusive
  $b \to s \gamma$ branching ratio}'',
  \href{http://dx.doi.org/10.1016/S0370-2693(98)00404-3}{{\em Phys. Lett.} {\bf
  B429} (1998)  169--187}.

\bibitem{Borzumati:1999qt}
F.~Borzumati, C.~Greub, T.~Hurth, and D.~Wyler, ``{Gluino contribution to
  radiative B decays: Organization of QCD corrections and leading order
  results}'', \href{http://dx.doi.org/10.1103/PhysRevD.62.075005}{{\em Phys.
  Rev.} {\bf D62} (2000)  075005},
  \href{http://arxiv.org/abs/hep-ph/9911245}{{\tt arXiv:hep-ph/9911245}}.

\bibitem{Greub:1999yq}
C.~Greub, T.~Hurth, and D.~Wyler, ``{Indirect search for supersymmetry in rare
  B decays}'', \href{http://arxiv.org/abs/hep-ph/9912420}{{\tt
  arXiv:hep-ph/9912420}}.

\bibitem{Wyler:2000mk}
D.~Wyler and F.~Borzumati, ``{Gluino contribution to radiative B decays: New
  operators, organization of QCD corrections and leading order results}'',
  \href{http://arxiv.org/abs/hep-ph/0104046}{{\tt arXiv:hep-ph/0104046}}.

\bibitem{Gambino:2001ew}
P.~Gambino and M.~Misiak, ``{Quark mass effects in $\bar B \to X_s \gamma$}'',
  \href{http://dx.doi.org/10.1016/S0550-3213(01)00347-9}{{\em Nucl. Phys.} {\bf
  B611} (2001)  338--366}, \href{http://arxiv.org/abs/hep-ph/0104034}{{\tt
  arXiv:hep-ph/0104034}}.

\bibitem{Hurth:2003dk}
T.~Hurth, E.~Lunghi, and W.~Porod, ``{Untagged $\bar B \to X_{s+d} \gamma$ CP
  asymmetry as a probe for new physics}'',
  \href{http://dx.doi.org/10.1016/j.nuclphysb.2004.10.024}{{\em Nucl. Phys.}
  {\bf B704} (2005)  56--74}, \href{http://arxiv.org/abs/hep-ph/0312260}{{\tt
  arXiv:hep-ph/0312260}}.

\bibitem{Ciuchini:2007cw}
M.~Ciuchini, E.~Franco, D.~Guadagnoli, V.~Lubicz, M.~Pierini, {\em et al.},
  ``{$D$ - $\bar{D}$ mixing and new physics: General considerations and
  constraints on the MSSM}'',
  \href{http://dx.doi.org/10.1016/j.physletb.2007.08.055}{{\em Phys. Lett.}
  {\bf B655} (2007)  162--166}, \href{http://arxiv.org/abs/hep-ph/0703204}{{\tt
  arXiv:hep-ph/0703204}}.

\bibitem{Paradisi:2005fk}
P.~Paradisi, ``{Constraints on SUSY lepton flavor violation by rare
  processes}'', \href{http://dx.doi.org/10.1088/1126-6708/2005/10/006}{{\em
  JHEP} {\bf 0510} (2005)  006},
  \href{http://arxiv.org/abs/hep-ph/0505046}{{\tt arXiv:hep-ph/0505046}}.

\bibitem{aga2}
K.~Agashe and M.~Graesser, ``{Supersymmetry breaking and the supersymmetric
  flavor problem: An Analysis of decoupling the first two generation
  scalars}'', \href{http://dx.doi.org/10.1103/PhysRevD.59.015007}{{\em Phys.
  Rev.} {\bf D59} (1999)  015007},
  \href{http://arxiv.org/abs/hep-ph/9801446}{{\tt arXiv:hep-ph/9801446}}.

\bibitem{Bagger:1997gg}
J.~A. Bagger, K.~T. Matchev, and R.-J. Zhang, ``{QCD corrections to
  flavor-changing neutral currents in the supersymmetric standard model}'',
  \href{http://dx.doi.org/10.1016/S0370-2693(97)00920-9}{{\em Phys. Lett.} {\bf
  B412} (1997)  77--85},
\href{http://arxiv.org/abs/hep-ph/9707225}{{\tt arXiv:hep-ph/9707225}}.
%%CITATION = HEP-PH/9707225;%%.

\bibitem{Ellis:2009di}
J.~Ellis, R.~N. Hodgkinson, J.~S. Lee, and A.~Pilaftsis, ``{Flavour Geometry
  and Effective Yukawa Couplings in the MSSM}'',
  \href{http://dx.doi.org/10.1007/JHEP02(2010)016}{{\em JHEP} {\bf 1002} (2010)
   016}, \href{http://arxiv.org/abs/0911.3611}{{\tt arXiv:0911.3611 [hep-ph]}}.

\bibitem{Blum:2009sk}
K.~Blum, Y.~Grossman, Y.~Nir, and G.~Perez, ``{Combining $K^0$ - $\bar K^0$
  Mixing and $D^0$~-~$\bar D^0$ Mixing to Constrain the Flavor Structure of New
  Physics}'', \href{http://dx.doi.org/10.1103/PhysRevLett.102.211802}{{\em
  Phys. Rev. Lett.} {\bf 102} (2009)  211802},
\href{http://arxiv.org/abs/0903.2118}{{\tt arXiv:0903.2118 [hep-ph]}}.
%%CITATION = 0903.2118;%%.

\end{thebibliography}
\end{document}